\documentclass[12pt]{iopart}
\usepackage{graphicx}% Include figure files
\usepackage{epstopdf}
\usepackage{amssymb}
\usepackage[ulem=normalem]{changes}

\begin{document}

\title[Exactly solvable models of nuclei]
{Exactly solvable models of nuclei}
\author{
P.~Van~Isacker$^1$ and K.~Heyde$^2$}

\address{
$^1$Grand Acc\'el\'erateur National d'Ions Lourds (GANIL), CEA/DSM--CNRS/IN2P3,
Boulevard Henri Becquerel, BP 55027, F-14076 Caen cedex 5, France}

\address{
$^2$Department of Physics and Astronomy, University of Ghent,
Proeftuinstraat 86, B-9000 Ghent, Belgium}

%\begin{abstract}
\bigskip

\noindent
``If politics is the art of the possible,
research is surely the art of the soluble~\cite{Medawar67}''
%\end{abstract}

\section{Introduction}
\label{s_intro}
The atomic nucleus is a many-body system
predominantly governed by a complex and effective in-medium nuclear interaction
and as such exhibits a rich spectrum of properties.
These range from independent nucleon motion in nuclei near closed shells,
to correlated two-nucleon pair formation
as well as collective effects characterized by vibrations and rotations
resulting from the cooperative motion of many nucleons.

The present-day theoretical description of the observed variety of nuclear excited states
has two possible microscopic approaches as its starting point.
Self-consistent mean-field methods start
from a given nucleon--nucleon effective force
or energy functional to construct the average nuclear field;
this leads to a description of collective modes
starting from the correlations between all neutrons and protons
constituting a given nucleus~\cite{Bender03}.
The spherical nuclear shell model, on the other hand,
includes all possible interactions between neutrons and protons
outside a certain closed-shell configuration~\cite{Caurier05}.
Both approaches make use of numerical algorithms
and are therefore computer intensive.

In this paper a review is given of a class of sub-models of both approaches,
characterized by the fact that they can be solved exactly,
highlighting in the process a number of generic results
related to both the nature of pair-correlated systems
as well as collective modes of motion in the atomic nucleus.
Exactly solvable models necessarily are of a schematic character,
valid for specific nuclei only.
But they can be used as a reference or `bench mark'
in the study of data over large regions of the nuclear chart
(series of isotopes or isotones)
with more realistic models using numerical approaches.
The emphasis here is on the exactly solvable models themselves
rather than on the comparison with data.
The latter aspect of exactly solvable models
is treated in several of the books mentioned at the end of this review
({\it e.g.}, references~\cite{Iachello87,Iachello91,Frank09,Rowe10}).

\section{An algebraic formulation of the quantal $n$-body problem}
\label{s_nbody}
Symmetry techniques and algebraic methods
are not confined to certain models in nuclear physics
but can be applied generally 
to find particular solutions of the quantal $n$-body problem.
How that comes about is explained in this section.

To describe the stationary properties
of an $n$-body system in non-relativistic quantum mechanics,
one needs to solve the {\sl time-independent Schr\"odinger equation}
which reads
\begin{equation}
\hat H\Psi(\xi_1,\dots,\xi_n)=
E\Psi(\xi_1,\dots,\xi_n),
\label{e_schroed}
\end{equation}
where $\hat H$ is the many-body {\sl hamiltonian}
\begin{equation}
\hat H=
\sum_{k=1}^n
\left(\frac{\hat p_k^2}{2m_k}+\hat V_1(\xi_k)\right)+
\sum_{k<l}
\hat V_2(\xi_k,\xi_l)+
\sum_{k<l<m}
\hat V_3(\xi_k,\xi_l,\xi_m)+\cdots,
\label{e_ham1}
\end{equation}
with $m_k$ the mass and $\hat p_k^2/2m_k$ the kinetic energy of particle $k$.
The particles can be {\sl bosons} or {\sl fermions}.
They may carry an intrinsic spin
and/or be characterized by other intrinsic variables
(such as isospin the projection of which distinguishes between a neutron and a proton).
These variables of particle $k$,
together with its position $\bar r_k$,
are collectively denoted by $\xi_k$.
Besides the kinetic energy and a possible external potential $\hat V_1(\xi_k)$,
the hamiltonian~(\ref{e_ham1}) contains terms
that represent two-, three- and possible higher-body interactions
$\hat V_2(\xi_k,\xi_l)$, $\hat V_3(\xi_k,\xi_l,\xi_m)$, \dots between the constituent particles.
The stationary properties of the $n$-body quantal system
are determined by solving the Schr\"odinger equation~(\ref{e_schroed})
with the additional constraint that the solution $\Psi(\xi_1,\dots,\xi_n)$
must be symmetric under exchange of bosons
and anti-symmetric under exchange of fermions.

The hamiltonian~(\ref{e_ham1}) can be written equivalently in {\sl second quantization}.
The one-body part of it describes a system of independent, non-interacting particles,
and defines a basis consisting of single-particle states $\phi_\alpha(\xi_k)$,
where $\alpha$ characterizes a stationary state in the potential $\hat V_1$.
In {\sl Dirac's notation} this single-particle state
can be written as $\langle\xi_k|\alpha\rangle$,
with $|\alpha\rangle$ a ket vector
that can be obtained by applying the creation operator $c_\alpha^\dag$ to the vacuum,
$|\alpha\rangle=c_\alpha^\dag|{\rm o}\rangle$.
The hermitian adjoint bra vector can be obtained likewise
by applying (to the left) the annihilation operator $c_\alpha$,
$\langle\alpha|=\langle{\rm o}|c_\alpha$.
A many-body state can now succinctly be written as
$|\alpha\beta\dots\rangle=c_\alpha^\dag c_\beta^\dag\dots|{\rm o}\rangle$,
and the Pauli principle is implicitly satisfied by requiring
that the creation and annihilation operators $c_\alpha^\dag$ and $c_\alpha$
obey either {\bf commutation relations} if the particles are bosons
or {\bf anti-commutation relations} if they are fermions, {\it viz.}
\begin{equation*}
[c_\alpha,c_\beta^\dag]=\delta_{\alpha\beta},
\quad
[c_\alpha,c_\beta]=[c_\alpha^\dag,c_\beta^\dag]=0,
\end{equation*}
or
\begin{equation*}
\{c_\alpha,c_\beta^\dag\}=\delta_{\alpha\beta},
\quad
\{c_\alpha,c_\beta\}=\{c_\alpha^\dag,c_\beta^\dag\}=0,
\end{equation*}
respectively.
With the preceding definitions,
the hamiltonian~(\ref{e_ham1}) can be rewritten as
\begin{equation}
\hat H=
\sum_\alpha\epsilon_\alpha c_\alpha^\dag c_\alpha+
\sum_{\alpha\beta\gamma\delta}
v_{\alpha\beta\gamma\delta}c_\alpha^\dag c_\beta^\dag c_\gamma c_\delta+\cdots,
\label{e_ham2}
\end{equation}
where $\epsilon_\alpha$ are coefficients
related to the one-body term in the hamiltonian~(\ref{e_ham1}),
$v_{\alpha\beta\gamma\delta}$ to the two-particle interaction,
and so on.
The summations are over complete sets of single-particle states,
which in most applications are infinite in number.
Even if the summations are restricted to a finite set of single-particle states,
the solution of the Schr\"odinger equation remains a formidable task,
owing to the exponential increase of the dimension of the Hilbert space of many-body states
with the numbers of particles and of available single-particle states.

A straightforward solution
of (the Schr\"odinger equation
associated with) the hamiltonian~(\ref{e_ham2})
is available only when the particles are non-interacting.
In that case the $n$-body problem reduces to $n$ one-body problems,
leading to $n$-particle eigenstates
that are {\bf Slater permanents} for bosons or {\bf Slater determinants} for fermions,
{\it i.e.}, eigenstates of the form
$c_{\alpha_1}^\dag\dots c_{\alpha_n}^\dag|{\rm o}\rangle$.
A Slater permanent or determinant is an important concept
that emanates from {\sl Hartree(-Fock) theory}.
Although correlations can be implicitly included
by way of an average potential or mean field,
two- and higher-particle interactions
are not explicitly treated in Hartree(-Fock) theory
but Slater permanents or determinants do provide a basis
in which the interactions between particles can be diagonalized.
The main obstacle that prevents one from doing such a diagonalization
is the dimension of the basis.
The question therefore arises whether interactions exist
that bypass the diagonalization and that can be treated analytically.

A strategy for solving with symmetry techniques
{\em particular classes} of the many-body hamiltonian~(\ref{e_ham2})
starts from the observation that it can be rewritten
in terms of the operators $\hat u_{\alpha\beta}\equiv c_\alpha^\dag c_\beta$.
The latter operators can be shown,
both for bosons and for fermions,
to obey the following {\em commutation} relations:
\begin{equation}
[\hat u_{\alpha\beta},\hat u_{\alpha'\beta'}]=
\hat u_{\alpha\beta'}\delta_{\alpha'\beta}-
\hat u_{\alpha'\beta}\delta_{\alpha\beta'},
\label{e_commutator}
\end{equation}
implying that the $\hat u_{\alpha\beta}$ 
generate the unitary Lie algebra ${\rm U}(\Omega)$,
with $\Omega$ the dimension of the single-particle basis.
[In the commutator~(\ref{e_commutator}) it is assumed
that all indices refer to either bosons or fermions.
The case of mixed systems of bosons and fermions
will be dealt with separately in subsection~\ref{ss_susy}.]
The algebra ${\rm U}(\Omega)$
is the {\bf dynamical algebra} $G_{\rm dyn}$ of the problem,
in the sense that the hamiltonian as well as other operators
can be expressed in terms of its generators.
It is not a true symmetry of the hamiltonian but a broken one.
The breaking of the symmetry associated with $G_{\rm dyn}$ 
is done in a particular way
which can be conveniently summarized
by a {\em chain of nested Lie algebras},
\begin{equation}
G_1\equiv G_{\rm dyn}\supset
G_2\supset\cdots\supset
G_s\equiv  G_{\rm sym},
\label{e_chain}
\end{equation}
where the last algebra $G_{\rm sym}$ in the chain
is the true-symmetry algebra,
whose generators commute with the hamiltonian.
For example, if the hamiltonian is rotationally invariant,
the symmetry algebra is the algebra of rotations in three dimensions,
$G_{\rm sym}={\rm SO}(3)$.

To appreciate the relevance of the classification~(\ref{e_chain})
in connection with the many-body hamiltonian~(\ref{e_ham2}),
note that to a particular chain of nested algebras
corresponds a class of hamiltonians
that can be written as a linear combination
of Casimir operators associated with the algebras in the chain,
\begin{equation}
\hat H_{\rm DS}=
\sum_{r=1}^s\sum_m\kappa_{rm}\hat C_m[G_r],
\label{e_hamds}
\end{equation}
where $\kappa_{rm}$ are arbitrary coefficients.
The $\hat C_m[G_r]$ are so-called {\bf Casimir operators} of the algebra $G_r$;
they are written as linear combinations of products of the generators of $G_r$, up to order $m$,
and satisfy the important property that they commute with all generators of $G_r$,
$[\hat C_m[G_r],\hat g]=0$ for all $\hat g\in G_r$.
The Casimir operators in~(\ref{e_hamds}) satisfy $[\hat C_m[G_r],\hat C_{m'}[G_{r'}]]=0$,
that is, they all commute with each other.
This property is evident from the fact
that for a chain of nested algebras
all elements of $G_r$ are in $G_{r'}$ or {\it vice versa}.
Hence, the hamiltonian~(\ref{e_hamds})
is written as a sum of commuting operators
and as a result its eigenstates
are labelled by the quantum numbers
associated with these operators.
Note that the condition
of the {\em nesting} of the algebras in~(\ref{e_chain})
is crucial for constructing a set of commuting operators
and hence for obtaining an analytic solution.
Casimir operators can be expressed
in terms of the operators $\hat u_{\alpha\beta}$
so that the expansion~(\ref{e_hamds})
can, in principle, be rewritten in the form~(\ref{e_ham2})
with the order of the interactions
determined by the maximal order $m$ of the invariants.

To summarize these results,
the hamiltonian~(\ref{e_hamds}),
which can be obtained from the general hamiltonian~(\ref{e_ham2})
for {\em specific choices} of the coefficients
$\epsilon_\alpha$, $\upsilon_{\alpha\beta\gamma\delta}$,\dots,
can be solved analytically.
Its eigenstates are characterized by quantum numbers $\Gamma_r$
which label {\sl irreducible representations} of the different algebras $G_r$
appearing in the reduction~(\ref{e_chain}),
leading to a classification that can conveniently be summarized as follows:
\begin{equation*}
\begin{array}{ccccccc}
G_1&\supset&G_2&\supset&\cdots&\supset&G_s\\
\downarrow&&\downarrow&&&&\downarrow\\
\Gamma_1&&\Gamma_2&&&&\Gamma_s
\end{array}.
\end{equation*}
The secular equation associated with the hamiltonian~(\ref{e_hamds})
is solved analytically
\begin{equation*}
\hat H_{\rm DS}
|\Gamma_1\Gamma_2\dots\Gamma_s\rangle=
\sum_{r=1}^s\sum_m\kappa_{rm}E_m(\Gamma_r)
|\Gamma_1\Gamma_2\dots\Gamma_s\rangle,
\end{equation*}
where $E_m(\Gamma_r)$ is the eigenvalue of the Casimir operator $\hat C_m[G_r]$
in the irreducible representation $\Gamma_r$.
The most important property of the hamiltonian~(\ref{e_hamds})
is that, while its energy eigenvalues
are known functions of the parameters $\kappa_{rm}$,
its eigenfunctions do not depend on $\kappa_{rm}$ and have a fixed structure.
Hamiltonians with the above properties
are said to have a {\bf dynamical symmetry}.
The symmetry $G_{\rm dyn}$ is broken
and the only remaining symmetry is $G_{\rm sym}$
which is the true symmetry of the problem.
This idea has found repeated and fruitful application
in many branches of physics,
and in particular in nuclear physics.

\section{The nuclear shell model}
\label{s_shell}
The basic structure of nuclei
can be derived from a few essential characteristics
of the nuclear mean field and the residual interaction.
A schematic hamiltonian that grasps the essential features
of nuclear many-body physics is of the form
\begin{equation}
\hat H=
\sum_{k=1}^A
\left(
{\frac{\hat p_k^2}{2m_k}}+
{\frac12}m_k\omega^2r_k^2+
\zeta_{\ell\ell}\,\hat\ell_k^2+
\zeta_{\ell s}\,\hat\ell_k\cdot\hat s_k
\right)+
\sum_{k<l}
\hat V_{\rm ri}(\xi_k,\xi_l),
\label{e_hamsm}
\end{equation}
where the indices $k,l$ run from 1 to $A$,
the number of nucleons in the nucleus.
The different terms in the hamiltonian~(\ref{e_hamsm})
are the kinetic energy,
a harmonic-oscillator potential with frequency $\omega$
(which is a first-order approximation to the nuclear mean field),
the quadratic orbital and spin--orbit terms,
and the residual two-nucleon interaction.

For a general residual interaction $\hat V_{\rm ri}(\xi_k,\xi_l)$
the hamiltonian~(\ref{e_hamsm}) must be solved numerically.
Two types of interaction lead to solvable models:
pairing (section~\ref{ss_racah}) and quadrupole (section~\ref{ss_elliott}).

\subsection{Racah's seniority model}
\label{ss_racah}
\begin{figure}
\centering
\includegraphics[width=10cm]{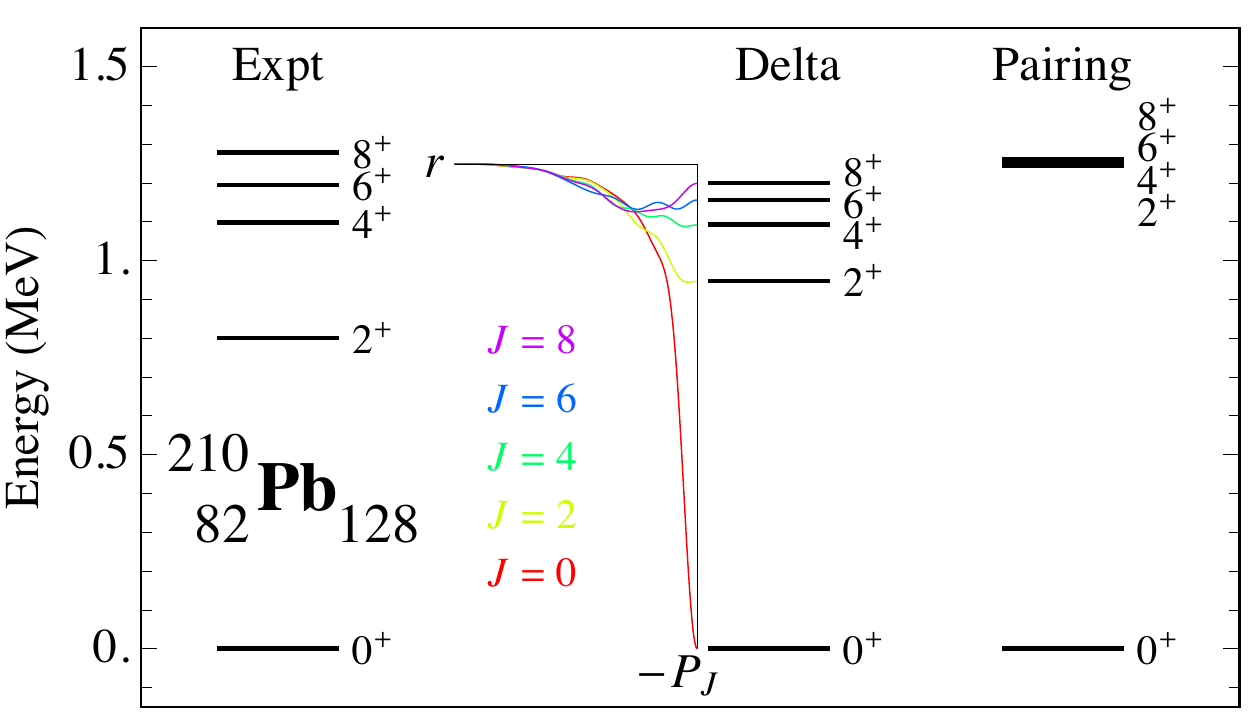}
\caption{
The experimental low-energy spectrum of $^{210}$Pb (left),
and the corresponding spectra for a zero-range delta (middle)
and for a pairing interaction (right).
Levels are labelled
by their angular momentum and parity $J^\pi$.
The inset shows the probability density $P_J$ to find
two nucleons at a distance $r$
when they are in the $2g_{9/2}$ orbit of a harmonic oscillator
and coupled to angular momentum $J$.}
\label{f_pb210}
\end{figure}

The nuclear force between identical nucleons
produces a large energy gap between $J=0$ and $J>0$ states,
and therefore can be approximated by a {\bf pairing} interaction
which only affects the ``paired'' $J=0$ state.
For nucleons in a single-$j$ shell,
pairing is defined by the two-body matrix elements
\begin{equation}
\langle j^2;JM_J|\hat V_{\rm pairing}|j^2;JM_J\rangle=
-{\frac12}g(2j+1)\delta_{J0}\delta_{M_J0},
\label{e_pairing}
\end{equation}
where $j$ is the orbital+spin angular momentum of a single nucleon
(hence $j$ is half-odd-integer),
$J$ results from the coupling of the angular momenta $j$ of the two nucleons
and $M_J$ is the projection of $J$ on the $z$ axis.
Furthermore, $g$ is the strength of the pairing interaction
which is attractive in nuclei ($g>0$).
Pairing is a reasonable, albeit schematic, approximation
to the residual interaction between {\em identical} nucleons
and hence can only be appropriate in semi-magic nuclei
with valence nucleons of a single type, either neutrons or protons.
The degree of approximation is illustrated in figure~\ref{f_pb210}
for the nucleus $^{210}$Pb
which can be described as two neutrons in the $1g_{9/2}$ orbit
outside the doubly magic $^{208}$Pb inert core.
Also shown is the probability density $P_J$ to find
two nucleons at a distance $r$
when they are in the $2g_{9/2}$ orbit of the harmonic oscillator
and coupled to angular momentum $J$.
This probability density at $r=0$ matches
the energies of the zero-range delta interaction.
The profiles of $P_J(r)$ for the different angular momenta
show that any attractive short-range interaction
favours the formation of a $J=0$ pair.
This basic property of the nuclear force is accounted for by pairing.

The pairing interaction was introduced by Racah
for the classification of electrons in an atom~\cite{Racah43}.
He was able to derive a closed formula
for the interaction energy among the electrons
and to prove that any eigenstate of the pairing interaction
is characterized by a `seniority number' $\upsilon$
which corresponds to the number of electrons
that are not in pairs coupled to orbital angular momentum $L=0$.
Racah's original definition of {\bf seniority}
made use of {\sl coefficients of fractional parentage}.
He later noted that simplifications arose
through the use of group theory~\cite{Racah49}.
Seniority turned out to be a label
associated with the (unitary) symplectic algebra ${\rm Sp}(2j+1)$
in the classification
\begin{equation}
\begin{array}{ccccc}
{\rm U}(2j+1)&\supset&{\rm Sp}(2j+1)&\supset&{\rm SU}(2)\\
\downarrow&&\downarrow&&\downarrow\\[0mm]
[1^n]&&[1^\upsilon]&&J
\end{array}.
\label{e_clasj}
\end{equation}
Since the nucleons are identical,
all states of the $j^n$ configuration belong to
the totally anti-symmetric irreducible representation $[1^n]$ of ${\rm U}(2j+1)$.
The irreducible representations of ${\rm Sp}(2j+1)$
therefore must also be totally anti-symmetric of the type $[1^\upsilon]$
with allowed values of seniority $\upsilon=n,n-2,\dots,1$ or 0.

In the definition~(\ref{e_clasj})
seniority appears as a label associated with the algebra ${\rm Sp}(2j+1)$.
This has the drawback that, depending on $j$,
the algebra can be quite large.
Matters become even more complicated
when the nucleons are non-identical
and have isospin $t={\frac12}$.
The total number of single-particle states is then $\Omega\equiv(2j+1)(2t+1)$
and one quickly runs into formidable group-theoretical reduction problems.
Fortunately, an alternative and simpler definition of seniority can be given
in terms of algebras that do not change with $j$.
The idea was simultaneously and independently proposed
by Kerman~\cite{Kerman61} for $t=0$ ({\it i.e.}, for identical nucleons)
and by Helmers~\cite{Helmers61} for general $t$.
It starts from operators $\hat S^j_+$ and $\hat S^j_-$
that create and annihilate {\em pairs} of particles in a single-$j$ shell
and the commutator of which leads to a third kind of generator, $\hat S^j_z$,
with one particle creation and one particle annihilation operator.
This set of operators, known as {\bf quasi-spin} operators, closes under commutation
and forms the (unitary) symplectic algebra ${\rm Sp}(4t+2)$
which can be shown to have equivalent properties to those of ${\rm Sp}(2j+1)$,
introduced in the classification~(\ref{e_clasj}).

The quasi-spin formulation of the pairing problem
relies on the fact that the pairing interaction
is related to the quadratic Casimir operator of the algebra ${\rm Sp}(4t+2)$.
This allows a succinct and simultaneous derivation of the eigenvalues
in the cases of identical nucleons ($t=0$)
and of neutrons and protons ($t={\frac12}$).
Over the years many results have been derived
and many extensions have been considered in both cases,
which are discussed separately in the following.

\subsubsection{Identical nucleons.}
\label{ss_racah0}
For $t=0$ one obtains the algebra Sp(2) which is {\sl isomorphic} to SU(2).
Due to its formal analogy with the spin algebra,
the name `quasi-spin' was coined by Kerman~\cite{Kerman61},
and this terminology has stuck for all cases, even when $t\neq0$.

The quasi-spin algebra ${\rm Sp}(2)\sim{\rm SU}(2)$
is obtained by noting that, in second quantization,
the pairing interaction defined in equation~(\ref{e_pairing}) is written as
\begin{equation}
\hat V_{\rm pairing}=
-g\hat S^j_+\hat S^j_-,
\label{e_pairing0}
\end{equation}
with
\begin{equation}
\hat S^j_+=
{\frac 1 2}\sqrt{2j+1}\,
(a_j^\dag\times a_j^\dag)^{(0)}_0,
\qquad
\hat S^j_-=\left(\hat S^j_+\right)^\dag,
\label{e_qspin0}
\end{equation}
where $a_{jm_j}^\dag$
creates a nucleon in orbit $j$ with projection $m_j$.
No isospin labels $t$ and $m_t$
are needed to characterize the identical nucleons.
The symbol $\times$ refers to {\sl coupling in angular momentum}
and $\hat S^j_+$ therefore creates a pair of nucleons
coupled to angular momentum $J=0$.
The commutator $[\hat S^j_+,\hat S^j_-]\equiv2\hat S^j_z$,
together with $[\hat S^j_z,\hat S^j_\pm]=\pm\hat S^j_\pm$,
shows that $\hat S^j_+$, $\hat S^j_-$ and $\hat S^j_z$
form a closed algebra SU(2).

Several emblematic results can be derived on the basis of SU(2).
The quasi-spin symmetry allows the determination
of the complete eigenspectrum of the pairing interaction
which is given by
\begin{equation}
\hat V_{\rm pairing}|j^n\upsilon JM_J\rangle=
E(n,\upsilon)|j^n\upsilon JM_J\rangle,
\label{e_spec0}
\end{equation}
with
\begin{equation}
E(n,\upsilon)=
-{\frac g4}(n-\upsilon)(2j-n-\upsilon+3).
\label{e_ener0}
\end{equation}
Besides the nucleon number $n$,
the total angular momentum $J$ and its projection $M_J$,
all eigenstates are characterized by a seniority quantum number $\upsilon$
which counts the {\em number of nucleons
not in pairs coupled to angular momentum zero}.
For an attractive pairing interaction ($g>0$),
the eigenstate with lowest energy has seniority $v=0$
if the nucleon number $n$ is even
and $v=1$ if $n$ is odd.
These lowest-energy eigenstates can,
up to a normalization factor, be written as
$(\hat S^j_+)^{n/2}|{\rm o}\rangle$ for even $n$
and $a_{jm_j}^\dag(\hat S^j_+)^{n/2}|{\rm o}\rangle$ for odd $n$,
where $|{\rm o}\rangle$ is the vacuum state for the nucleons.

The discussion of pairing correlations in nuclei traditionally has been inspired
by the treatment of {\sl superfluidity} in condensed matter,
explained in 1957 by Bardeen, Cooper and Schrieffer~\cite{BCS57},
and later adapted to the discussion of pairing in nuclei~\cite{Bohr58}.
The superfluid phase is characterized
by the presence of a large number of identical bosons in a single quantum state.
In superconductors the bosons are pairs of electrons
with opposite momenta that form at the Fermi surface
while in nuclei, according to the preceding discussion,
they are pairs of valence nucleons with opposite angular momenta.

A generalization of these concepts
concerns that towards several orbits.
In case of degenerate orbits
this can be achieved by making the substitution
$\hat S^j_\mu\mapsto\hat S_\mu\equiv\sum_j\hat S^j_\mu$
which leaves all preceding results, valid for a single-$j$ shell, unchanged.
The ensuing formalism
can then be applied to semi-magic nuclei
but, since it requires the assumption
of a pairing interaction with degenerate orbits,
its applicability is limited.

\begin{figure}
\centering
\includegraphics[width=10cm]{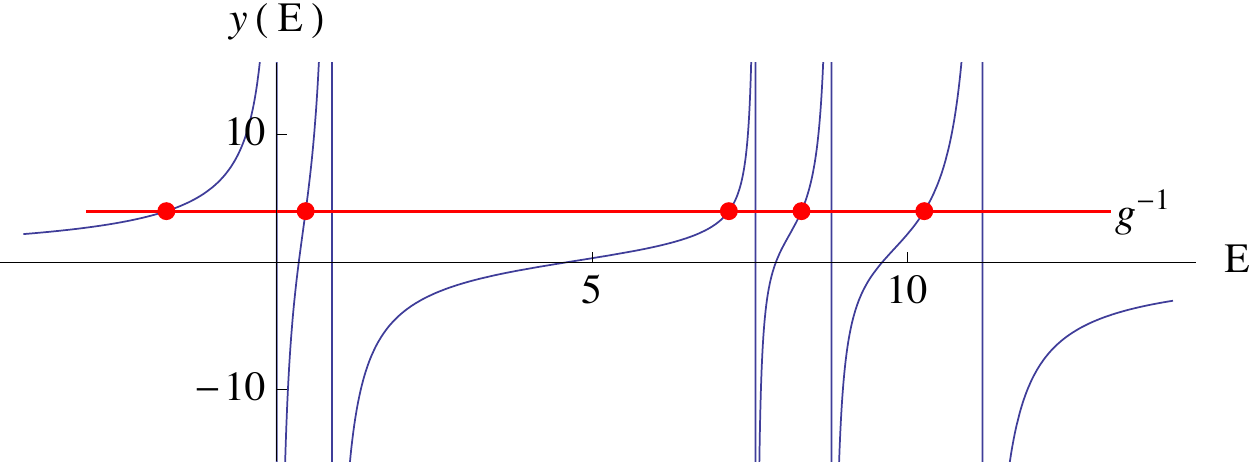}
\caption{
Graphical solution of the Richardson equation for two nucleons
distributed over five single-particle orbits.
The sum $\sum_j\Omega_j/(2\epsilon_j-E)\equiv y(E)$ (in MeV$^{-1}$)
is plotted as a function of $E$ (in MeV).
The intersections (red dots) of this curve (blue) with the (red) line $y=1/g$
correspond to the solutions of the Richardson equation.}
\label{f_richardson}
\end{figure}
An exact method to solve the problem of particles
distributed over non-degenerate levels
interacting through a pairing force
was proposed by Richardson~\cite{Richardson63} based on the Bethe {\it ansatz}
and has been generalized more recently to other classes
of integrable pairing models~\cite{Dukelsky01}.
Richardson's approach can be illustrated
by supplementing the pairing interaction~(\ref{e_pairing0})
with non-degenerate single-particle energies,
to obtain the following hamiltonian:
\begin{equation}
\hat H_{\rm pairing}=
\sum_j\epsilon_j\hat n_j-g\hat S_+\hat S_-,
\label{e_richardson1}
\end{equation}
where $\hat n_j$ is the number operator for orbit $j$,
$\epsilon_j$ is the single-particle energy of that orbit
and $\hat S_\pm=\sum_j\hat S^j_\pm$.
The solvability of the hamiltonian~(\ref{e_richardson1})
arises as a result of the symmetry
${\rm SU}(2)\otimes{\rm SU}(2)\otimes\cdots$
where each SU(2) algebra pertains to a specific $j$.
The eigenstates are of the form
\begin{equation}
\prod_{p=1}^{n/2}
\left(\sum_j{\frac{\hat S^j_+}{2\epsilon_j-E_p}}\right)
|{\rm o}\rangle,
\label{e_ansatz}
\end{equation}
where the $E_p$ are solutions of $n/2$ coupled, non-linear Richardson equations~\cite{Richardson63}
\begin{equation}
\sum_j{\frac{\Omega_j}{2\epsilon_j-E_p}}-
\sum_{p'(\neq p)}^{n/2}{\frac{2}{E_{p'}-E_p}}=
{\frac 1 g},
\qquad
p=1,\dots,n/2,
\label{e_richeqs}
\end{equation}
with $\Omega_j=j+1/2$.
This equation is solved graphically
for the simple case of $n=2$ in figure~\ref{f_richardson}.
Each pair in the product~(\ref{e_ansatz}) is defined
through coefficients $\alpha_j=(2\epsilon_j-E_p)^{-1}$
which depend on the energy $E_p$
where $p$ labels the $n/2$ pairs.   
A characteristic feature of the Bethe {\it ansatz} is
that it no longer consists of a superposition of {\em identical} pairs
since the coefficients $(2\epsilon_j-E_p)^{-1}$
vary as $p$ runs from 1 to $n/2$.
Richardson's model thus provides a solution
that covers all possible hamiltonians~(\ref{e_richardson1}),
ranging from those with superfluid character
to those with little or no pairing correlations.
Whether the solution can be called superfluid 
depends on the differences $\epsilon_j-\epsilon_{j'}$
in relation to the strength $g$.

The pairing hamiltonian~(\ref{e_richardson1})
admits non-degenerate single-particle orbits $\epsilon_j$
but requires a constant strength $g$ of the pairing interaction,
independent of $j$.
Alternatively, a hamiltonian with degenerate single-particle orbits $\epsilon_j=\epsilon$
but orbit-dependent strengths $g_j$,
\begin{equation}
\hat H'_{\rm pairing}=
\epsilon\sum_j\hat n_j-\sum_jg_j\hat S^j_+\sum_{j'}g_{j'}\hat S^{j'}_-,
\label{e_richardson2}
\end{equation}
can also be solved exactly based on the Bethe {\it ansatz}~\cite{Pan98}.
No exact solution is known, however,
of a pairing hamiltonian with non-degenerate single-particle orbits $\epsilon_j$
and orbit-dependent strengths $g_j$,
except in the case of two orbits~\cite{Balantekin07}.
Solvability by Richardson's technique requires the pairing interaction
to be separable with strengths that satisfy $g_{jj'}=g_j g_{j'}$
and no exact solution is known in the non-separable case when $g_{jj'}\neq g_j g_{j'}$.

These possible generalizations notwithstanding,
it should be kept in mind that a pairing interaction
is but an approximation to a realistic residual interaction among nucleons,
as is clear from figure~\ref{f_pb210}.
A more generally valid approach is obtained
if one imposes the following condition
on the shell-model hamiltonian~(\ref{e_hamsm}):
\begin{equation}
[[\hat H_{\rm GS},\hat S^\alpha_+],\hat S^\alpha_+]=
\Delta\left(\hat S^\alpha_+\right)^2,
\label{e_gscon}
\end{equation}
where $\Delta$ is a constant
and $\hat S^\alpha_+=\sum_j\alpha_jS^j_+$
creates the lowest two-particle eigenstate of $\hat H_{\rm GS}$
with energy $E_0$,
$\hat H_{\rm GS}\hat S^\alpha_+|{\rm o}\rangle=E_0\hat S^\alpha_+|{\rm o}\rangle$.
The condition~(\ref{e_gscon}) of {\bf generalized seniority}, proposed by Talmi~\cite{Talmi71},
is much weaker than the assumption of a pairing interaction
and it does not require
that the commutator $[\hat S^\alpha_+,\hat S^\alpha_-]$
yields (up to a constant) the number operator
which is central to the quasi-spin formalism.
In spite of the absence of a closed algebraic structure,
it is still possible to compute exact results
for hamiltonians satisfying the condition~(\ref{e_gscon}).
For an even number of nucleons, its ground state
has the same simple structure as in the quasi-spin formalism,
\begin{equation*}
\hat H_{\rm GS}\left(\hat S^\alpha_+\right)^{n/2}|{\rm o}\rangle=
E_{\rm GS}(n)\left(\hat S^\alpha_+\right)^{n/2}|{\rm o}\rangle,
\end{equation*}
with an energy that can be computed for any nucleon number $n$,
\begin{equation*}
E_{\rm GS}(n)=nE_0+{\frac12}n(n-1)\Delta.
\end{equation*}
Because of its linear and quadratic dependence on the nucleon number $n$,
this result can be considered
as a generalization of Racah's seniority formula~(\ref{e_ener0}),
to which it reduces if $E_0=-g(j+1)/2$ and $\Delta=g/2$.

\subsubsection{Neutrons and protons.}
\label{ss_racah1}
For $t={\frac12}$ one obtains the quasi-spin algebra Sp(4)
which is {\sl isomorphic} to SO(5).
The algebra ${\rm Sp}(4)$ or ${\rm SO}(5)$ is characterized by two labels,
corresponding to {\bf seniority} $\upsilon$ and {\bf reduced isospin} $T_\upsilon$.
Seniority $\upsilon$ has the same interpretation as in the like-nucleon case,
namely the number of nucleons not in pairs coupled to angular momentum $J=0$,
while reduced isospin $T_\upsilon$
corresponds to the total isospin of these nucleons~\cite{Flowers52,Racah52}.

The above results are obtained from the general analysis
as carried out by Helmers~\cite{Helmers61} for any $t$.
It is of interest to carry out the analysis explicitly
for the choice which applies to nuclei, namely $t={\frac12}$.
Results are given in $LS$ coupling,
which turns out to be the more convenient scheme
for the generalization to neutrons and protons.

\begin{figure}
\centering
\includegraphics[width=10cm]{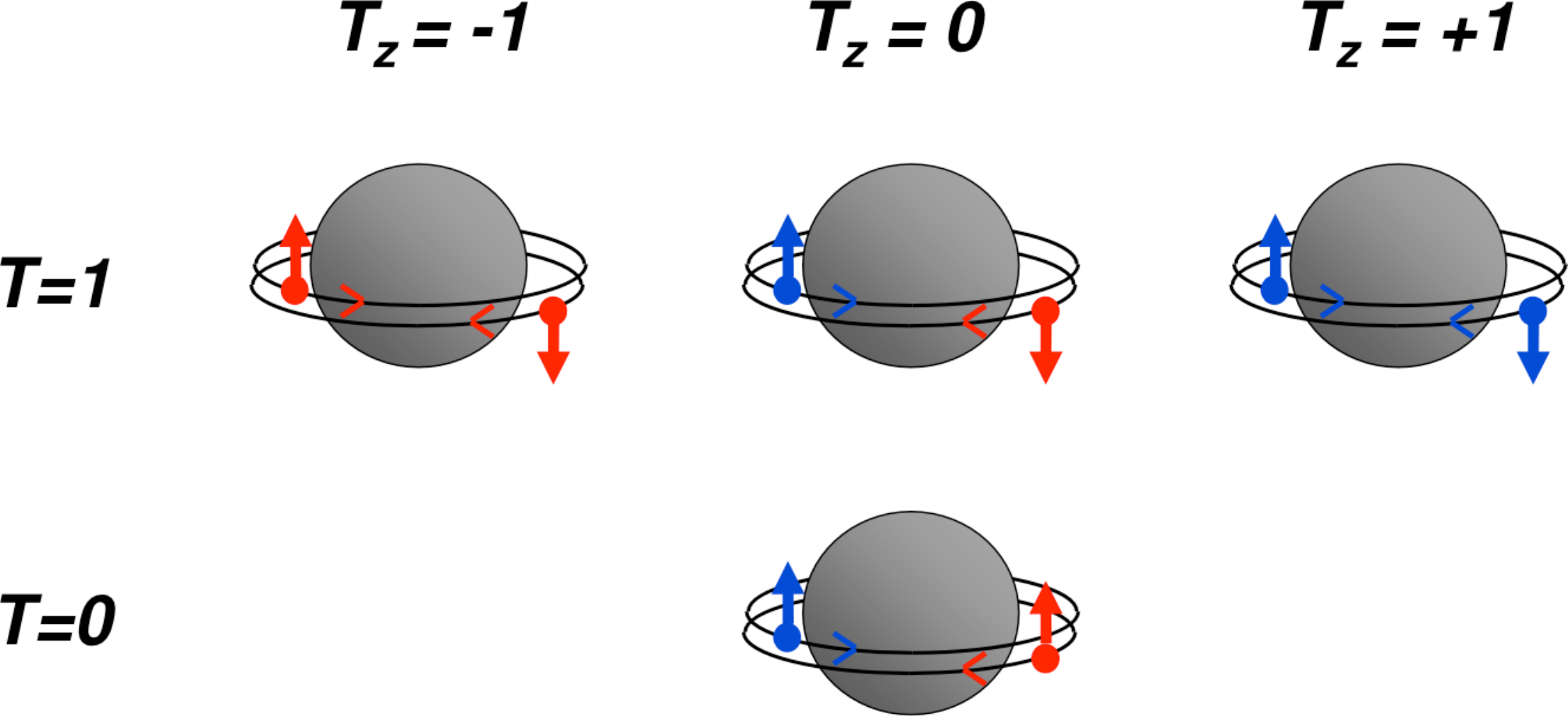}
\caption{
Schematic illustration of the different types of nucleon pairs
with orbital angular momentum $L=0$.
The valence neutrons (blue) or protons (red)
that form the pair occupy time-reversed orbits
(circling the nucleus in opposite direction).
If the nucleons are identical they must have anti-parallel spins---a
configuration which is also allowed for a neutron--proton pair (top).
The configuration with parallel spins
is only allowed for a neutron--proton pair (bottom).
Taken from reference~\cite{Frank09}.}
\label{f_pairs}
\end{figure}
If the $\ell$ shell contains neutrons and protons,
the pairing interaction is assumed to be isospin invariant,
which implies that it is the same
in the three possible $T=1$ channels,
neutron--neutron, neutron--proton and proton--proton,
and that the pairing interaction~(\ref{e_pairing0}) takes the form
\begin{equation}
\hat V'_{\rm pairing}=
-g\sum_\mu\hat S^\ell_{+,\mu}\hat S^\ell_{-,\mu}\equiv
-g\hat S^\ell_+\cdot\hat S^\ell_-,
\label{e_pairing2}
\end{equation}
where the dot indicates a scalar product in isospin.
In terms of the nucleon operators $a^\dag_{\ell m_\ell,sm_s,tm_t}$,
which now carry also isospin indices (with $t={\frac12}$),
the pair operators are
\begin{equation}
\hat S^\ell_{+,\mu}=
\sqrt{\frac12}\sqrt{2\ell+1}
(a_{\ell,s,t}^\dag\times a_{\ell,s,t}^\dag)^{(001)}_{00\mu},
\qquad
\hat S^\ell_{-,\mu}=
\left(\hat S^\ell_{+,\mu}\right)^\dag,
\label{e_qspin2}
\end{equation}
where $\hat S$ refers to a pair
with orbital angular momentum $L=0$, spin $S=0$ and isospin $T=1$.
The index $\mu$ (isospin projection) distinguishes
neutron--neutron ($\mu=+1$),
neutron--proton ($\mu=0$)
and proton--proton ($\mu=-1$) pairs.
There are thus three different pairs with $L=0$, $S=0$ and $T=1$
(top line in figure~\ref{f_pairs})
and they are related through the action
of the isospin raising and lowering operators $\hat T_\pm$.
The quasi-spin algebra
associated with the hamiltonian~(\ref{e_pairing2}) is SO(5)
and makes the problem analytically solvable~\cite{Hecht65}.

For a neutron and a proton there exists
a different paired state with {\em parallel} spins
(bottom line of figure~\ref{f_pairs}).
The most general pairing interaction for a system of neutrons and protons is therefore
\begin{equation}
\hat V''_{\rm pairing}=
-g\hat S^\ell_+\cdot\hat S^\ell_-
-g'\hat P^\ell_+\cdot\hat  P^\ell_-,
\label{e_pairing3}
\end{equation}
where $\hat P$ refers to a pair
with orbital angular momentum $L=0$, spin $S=1$ and isospin $T=0$,
\begin{equation}
\hat P^\ell_{+,\mu}=
\sqrt{\frac12}\sqrt{2\ell+1}
(a_{\ell,s,t}^\dag\times a_{\ell,s,t}^\dag)^{(010)}_{0\mu0},
\qquad
\hat P^\ell_{-,\mu}=
\left(\hat P^\ell_{+,\mu}\right)^\dag.
\label{e_qspin3}
\end{equation}
The index $\mu$ is the spin projection
and distinguishes the three spatial orientations of the $S=1$ pair.
The pairing interaction~(\ref{e_pairing3})
now involves two parameters $g$ and $g'$,
the strengths of the {\bf isovector} and {\bf isoscalar} components.
Solutions with an intrinsically different structure
are obtained for different ratios $g/g'$.

In general, the eigenproblem
associated with the pairing interaction~(\ref{e_pairing3})
can only be solved numerically
which, given a typical size of a shell-model space,
can be a formidable task.
However, for specific choices of $g$ and $g'$
the solution of $\hat V''_{\rm pairing}$
can be obtained analytically~\cite{Flowers64,Pang69}.
The analysis reveals the existence of a quasi-spin algebra SO(8)
formed by the pair operators~(\ref{e_qspin2}) and~(\ref{e_qspin3}),
their commutators, the commutators of these among themselves,
and so on until a closed algebraic structure is attained.
Closure is obtained by introducing,
in addition to the pair operators~(\ref{e_qspin2}) and~(\ref{e_qspin3}),
the number operator $\hat n$,
the spin and isospin operators $\hat S_\mu$ and $\hat T_\mu$,
and the Gamow-Teller-like operators $\hat U_{\mu\nu}$,
defined in section~\ref{ss_wigner}
in the context of Wigner's supermultiplet algebra.

From a study of the subalgebras of SO(8)
it can be concluded that the pairing interaction~(\ref{e_pairing3})
has a dynamical symmetry (in the sense of section~\ref{s_nbody})
in one of the three following cases:
(i) $g=0$, (ii) $g'=0$ and (iii) $g=g'$,
corresponding to pure isoscalar pairing, pure isovector pairing
and pairing with equal isoscalar and isovector strengths, respectively.
Seniority $\upsilon$ turns out to be conserved in these three limits
and associated with either an SO(5) algebra in cases (i) and (ii),
or with the SO(8) algebra in case (iii).

One of the main results of the theory of pairing between identical nucleons
is the recognition of the special structure of low-energy states
in terms of $S$ pairs.
It is therefore of interest to address the same question
in the theory of pairing between neutrons and protons.
The nature of SO(8) superfluidity can be illustrated
with the example of the ground state of nuclei
with an equal number of neutrons $N$ and protons $Z$.
For equal strengths of isoscalar and isovector pairing, $g=g'$,
the pairing interaction~(\ref{e_pairing3}) is solvable
and its ground state can be shown to be~\cite{Dobes98}:
\begin{equation}
\left(\hat S^\ell_+\cdot \hat S^\ell_+-
\hat P^\ell_+\cdot \hat P^\ell_+\right)^{n/4}
|{\rm o}\rangle.
\label{e_cond1}
\end{equation}
This shows that the superfluid solution
acquires a {\bf quartet} structure
in the sense that it reduces to a {\sl condensate} of a boson-like object,
which corresponds to four nucleons.
Since this object in~(\ref{e_cond1}) is scalar in spin and isospin,
it can be thought of as an $\alpha$ particle;
its orbital character, however, might be different
from that of an actual $\alpha$ particle.
A quartet structure is also present
in the other two limits of SO(8), with either $g=0$ or $g'=0$,
which have a ground-state wave function of the type~(\ref{e_cond1})
with either the first or the second term suppressed.
Thus, a reasonable {\it ansatz}
for the ground-state wave function of an $N=Z$ nucleus
of the pairing interaction~(\ref{e_pairing3})
with arbitrary strengths $g$ and $g'$ is
\begin{equation}
\left(\cos\theta\;\hat S^\ell_+\cdot \hat S^\ell_+-
\sin\theta\;\hat P^\ell_+\cdot \hat P^\ell_+\right)^{n/4}
|{\rm o}\rangle,
\label{e_cond2}
\end{equation}
where $\theta$ is a parameter
that depends on the ratio $g/g'$.

The condensate~(\ref{e_cond2}) of $\alpha$-like particles
can serve as a good approximation to the $N=Z$ ground state
of the pairing interaction~(\ref{e_pairing3})
for any combination of $g$ and $g'$~\cite{Dobes98}.
Nevertheless, it should be stressed that,
in the presence of both neutrons and protons in the valence shell,
the pairing interaction~(\ref{e_pairing3})
is {\em not} a good approximation to a realistic shell-model hamiltonian
which contains an important quadrupole component
(see, {\it e.g.}, the shell-model review~\cite{Caurier05}).
Consequently, any model based on $L=0$ fermion pairs only,
remains necessarily schematic in nature.
A realistic model should include also $L\neq0$ pairs.

\subsection{Wigner's supermultiplet model}
\label{ss_wigner}
Wigner's supermultiplet model~\cite{Wigner37}
assumes nuclear forces to be invariant under rotations
in {\em spin} as well as {\em isospin} space.
A shell-model hamiltonian with this property
satifies the following commutation relations:
\begin{equation}
[\hat H,\hat S_\mu]=
[\hat H,\hat T_\mu]=
[\hat H,\hat U_{\mu\nu}]=0,
\label{e_su4inv}
\end{equation}
where
\begin{equation}
\hat S_\mu=\sum_{k=1}^A\hat s_{k,\mu},
\qquad
\hat T_\mu=\sum_{k=1}^A\hat t_{k,\mu},
\qquad
\hat U_{\mu\nu}=\sum_{k=1}^A\hat s_{k,\mu}\hat t_{k,\nu},
\label{e_su4gen}
\end{equation}
are the spin, isospin and spin--isospin operators,
in terms of $\hat s_{k,\mu}$ and $\hat t_{k,\mu}$,
the spin and isospin components of nucleon $k$.
The 15 operators~(\ref{e_su4gen}) generate the Lie algebra SU(4).
According to the discussion in section~\ref{s_nbody},
any hamiltonian satisfying the conditions~(\ref{e_su4inv})
has SU(4) symmetry,
and this in addition to symmetries associated
with the conservation of total spin $S$ and total isospin $T$.

The physical relevance of Wigner's supermultiplet classification
is due to the short-range attractive nature
of the residual interaction
as a result of which states with spatial symmetry are favoured energetically.
To obtain a qualitative understanding of SU(4) symmetry,
it is instructive to analyze the case of two nucleons.
Total anti-symmetry of the wave function requires
that the spatial part is symmetric
and the spin-isospin part anti-symmetric or {\it vice versa}.
Both cases correspond to a different symmetry under SU(4),
the first being anti-symmetric and the second symmetric.
The symmetry under a given algebra 
can characterized by the so-called {\sl Young tableau}~\cite{Hamermesh62}.
For two nucleons the symmetric and anti-symmetric
irreducible representations are denoted by
\begin{equation*}
\Box\Box\equiv[2,0],
\qquad
\begin{array}{c}
\Box\\[-1ex]
\Box
\end{array}\equiv[1,1],
\end{equation*}
respectively, and the Young tableaux are conjugate,
that is, one is obtained from the other by interchanging rows and columns.
This result can be generalized to many nucleons,
leading to the conclusion that the energy of a state
depends on its SU(4) labels,
which are three in number and denoted here as
$(\bar\lambda,\bar\mu,\bar\nu)$.

Wigner's supermultiplet model is an $LS$-coupling scheme
which is not appropriate for nuclei.
In spite of its limited applicability,
Wigner's idea remains important
because it demonstrates the connection
between the short-range character of the residual interaction
and the spatial symmetry of the many-body wave function.
The break down of SU(4) symmetry
is a consequence of the spin--orbit term
in the shell-model hamiltonian~(\ref{e_hamsm})
which does not satisfy
the first and third commutator in equation~(\ref{e_su4inv}).
The spin--orbit term breaks SU(4) symmetry
[SU(4) irreducible representations are admixed by it]
and does so increasingly in heavier nuclei
since the energy splitting
of the spin doublets $\ell-{\frac12}$ and $\ell+{\frac12}$
increases with nucleon number $A$.
In addition, SU(4) symmetry
is also broken by the Coulomb interaction---an
effect that also increases with $A$---and
by spin-dependent residual interactions.

\subsection{Elliott's rotation model}
\label{ss_elliott}
In Wigner's supermultiplet model
the spatial part of the wave function 
is characterized by a total orbital angular momentum $L$
but is left unspecified otherwise.
The main feature of Elliott's model~\cite{Elliott58} is
that it provides additional orbital quantum numbers
that are relevant for {\sl deformed nuclei}.
Elliott's model of rotation presupposes Wigner's SU(4) classification
and assumes in addition that the residual interaction
has a {\bf quadrupole} character
which is a reasonable hypothesis
if the valence shell contains neutrons and protons.
One requires that the schematic shell-model hamiltonian~(\ref{e_hamsm}) reduces to
\begin{equation}
\hat H_{\rm SU(3)}=
\sum_{k=1}^A
\left(
{\frac{\hat p_k^2}{2m_k}}+
{\frac12}m_k\omega^2r_k^2
\right)+
\hat V_{\rm quadrupole},
\label{e_su3}
\end{equation}
where
$\hat V_{\rm quadrupole}=-g_2\hat Q\cdot\hat Q$
contains a quadrupole operator
\begin{equation}
\hat Q_\mu=
\sqrt{\frac32}
\left[
\sum_{k=1}^A\frac{1}{b^2}
(\bar r_k\wedge\bar r_k)^{(2)}_\mu+
\frac{b^2}{\hbar^2}
\sum_{k=1}^A(\bar p_k\wedge\bar p_k)^{(2)}_\mu
\right],
\label{e_su3q}
\end{equation}
in terms of coordinates $\bar r_k$ and momenta $\bar p_k$ of nucleon $k$,
and where $b$ is the oscillator length parameter,
$b=\sqrt{\hbar/m_{\rm n}\omega}$ with $m_{\rm n}$ the mass of the nucleon.

With use of the techniques explained in section~\ref{s_nbody},
it can be shown that the shell-model hamiltonian~(\ref{e_su3})
is analytically solvable.
Since the hamiltonian~(\ref{e_su3})
satisfies the commutation relations~(\ref{e_su4inv}),
it has SU(4) symmetry
and its eigenstates are characterized by the associated quantum numbers,
the supermultiplet labels $(\bar\lambda,\bar\mu,\bar\nu)$.
The spin--isospin symmetry SU(4) is equivalent through conjugation
to the orbital symmetry ${\rm U}(\Omega)$,
where $\Omega$ denotes the orbital shell size
({\it i.e.}, $\Omega=1,3,6,\dots$ for the $s$, $p$, $sd$,\dots shells).
The algebra ${\rm U}(\Omega)$, however,
is {\em not} a true symmetry of the hamiltonian~(\ref{e_su3})
but is broken according to the nested chain of algebras
${\rm U}(\Omega)\supset{\rm SU}(3)\supset{\rm SO}(3)$.
As a result one finds that the hamiltonian~(\ref{e_su3}) has the eigenstates 
$|[1^n](\bar\lambda,\bar\mu,\bar\nu)(\lambda,\mu)K_LLM_LSM_STM_T\rangle$
with energies
\begin{equation*}
E_{\rm SU(3)}(\lambda,\mu,L)=E_0-
g_2\left[
4(\lambda^2+\mu^2+\lambda\mu+3\lambda+3\mu)-3L(L+1)
\right],
\end{equation*}
where $E_0$ is a constant energy
associated with the first term in the hamiltonian~(\ref{e_su3}).
Besides the set of quantum numbers
encountered in Wigner's supermultiplet model,
that is, the SU(4) labels $(\bar\lambda,\bar\mu,\bar\nu)$,
the total orbital angular momentum $L$ and its projection $M_L$,
the total spin $S$ and its projection $M_S$,
and the total isospin $T$ and its projection $M_T$,
all eigenstates of the hamiltonian~(\ref{e_su3}) are characterized
by the SU(3) quantum numbers $(\lambda,\mu)$
and an additional label $K_L$.
Each irreducible representation $(\lambda,\mu)$
contains the orbital angular momenta $L$
typical of a {\sl rotational band},
cut off at some upper limit~\cite{Elliott58}.
The label $K_L$ defines the {\sl intrinsic state} associated to that band
and can be interpreted as the projection of the orbital angular momentum $L$
on the axis of symmetry of the rotating deformed nucleus.

The importance of Elliott's idea
is that it gives rise to a rotational classification of states
through mixing of spherical configurations.
With the SU(3) model it was shown, for the first time,
how deformed nuclear shapes
may arise out of the spherical shell model.
As a consequence, Elliott's work bridged the gap
between the spherical nuclear shell model
and the geometric collective model (see section~\ref{s_gcm})
which up to that time (1958) existed as separate views of the nucleus.

\begin{figure}
\centering
\includegraphics[width=10cm]{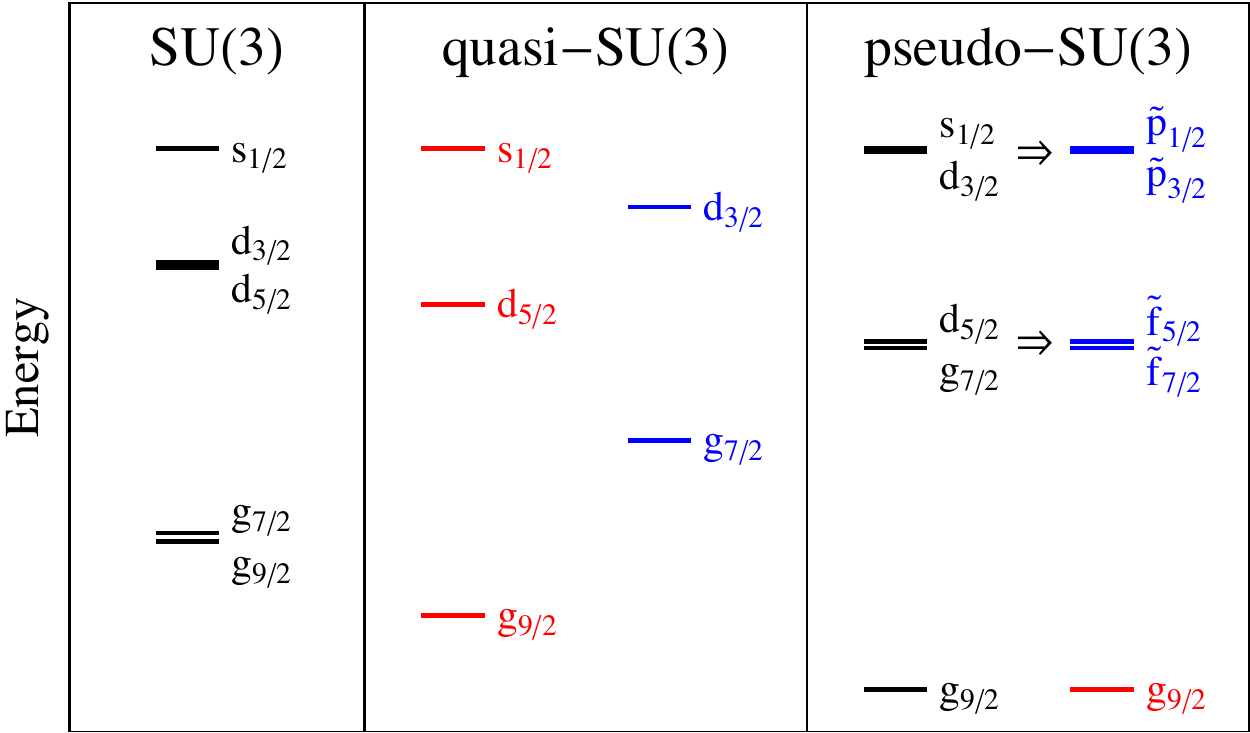}
\caption{The single-particle energies
(for a non-zero quadratic orbital strength, $\zeta_{\ell\ell}\neq0$)
in SU(3), quasi-SU(3) and pseudo-SU(3) for the example of the $sdg$ oscillator shell.
The spin--orbit strength
is $\zeta_{\ell s}\approx0$ in SU(3),
$\zeta_{\ell s}\approx2\zeta_{\ell\ell}$ in quasi-SU(3) 
and $\zeta_{\ell s}\approx4\zeta_{\ell\ell}$ in pseudo-SU(3).
The single-particle spaces in red and in blue
are assumed to be approximately decoupled.
In pseudo-SU(3) the level degeneracies
can be interpreted in terms of a pseudo-spin symmetry.}
\label{f_qpsu3}
\end{figure}
Elliott's SU(3) model provides a natural explanation
of rotational phenomena, ubiquitous in nuclei,
but it does so by assuming Wigner's SU(4) symmetry
which is known to be badly broken in most nuclei.
This puzzle has motivated much work since Elliott:
How can rotational phenomena in nuclei be understood
starting from a $jj$-coupling scheme which applies to most nuclei?
Over the years several schemes have been proposed
with the aim of transposing the SU(3) scheme
to those modified situations.
One such modification has been suggested by Zuker~{\it et al.}~\cite{Zuker95}
under the name of quasi-SU(3)
and it invokes the similarities of matrix elements of the quadrupole operator
in the $jj$- and $LS$-coupling schemes. 

Arguably the most successful way
to extend the applications of the SU(3) model to heavy nuclei
is based upon the concept of pseudo-spin symmetry.
The starting point for the explanation of this symmetry
is the single-particle part of the hamiltonian~(\ref{e_hamsm}).
For $\zeta_{\ell\ell}=\zeta_{\ell s}=0$
a three-dimensional isotropic harmonic oscillator is obtained
which exhibits degeneracies associated with U(3) symmetry.
For arbitrary non-zero values of $\zeta_{\ell\ell}$ and $\zeta_{\ell s}$
this symmetry is broken.
However, for the particular combination $4\zeta_{\ell\ell}=\zeta_{\ell s}$
some degree of degeneracy,
associated with a so-called pseudo-spin symmetry,
is restored in the single-particle spectrum (see figure~\ref{f_qpsu3}).

Pseudo-spin symmetry has a long history in nuclear physics.
The existence of nearly degenerate pseudo-spin doublets
in the nuclear mean-field potential
was pointed out almost forty years ago
by Hecht and Adler~\cite{Hecht69}
and by Arima~{\it et al.}~\cite{Arima69}
who noted that,
because of the small pseudo-spin--orbit splitting,
pseudo-$LS$ coupling should be a reasonable starting point
in medium-mass and heavy nuclei
where $LS$ coupling becomes unacceptable.
With pseudo-$LS$ coupling as a premise,
a pseudo-SU(3) model can be constructed~\cite{Ratna73}
in much the same way as Elliott's SU(3) model
can be defined in $LS$ coupling.
It is only many years after its original suggestion
that Ginocchio showed pseudo-spin
to be a symmetry of the Dirac equation
which occurs if the scalar and vector potentials
are equal in size but opposite in sign~\cite{Ginocchio97}.

The models discussed so far
all share the property of being confined to a single shell,
either an oscillator or a pseudo-oscillator shell.
A full description of nuclear collective motion requires correlations
that involve configurations outside a single (pseudo) oscillator shell.
The proper framework for such correlations
invokes the concept of a {\bf non-compact} algebra
which, in contrast to a compact one,
can have infinite-dimensional unitary irreducible representations.
The latter condition is necessary
since the excitations into higher shells
can be infinite in number.
The inclusion of excitations
into higher shells of the harmonic oscillator,
was achieved by Rosensteel and Rowe
by embedding the SU(3) algebra
into the (non-compact) symplectic algebra Sp(3,R)~\cite{Rosensteel77}.

\subsection{The Lipkin model}
\label{ss_lipkin}
Another noteworthy algebraic model in nuclear physics
is due to Lipkin {\it et al.}~\cite{Lipkin65}
who consider two levels (assigned an index $\sigma=\pm$)
each with degeneracy $\Omega$
over which $n$ fermions are distributed.
The Lipkin model has an SU(2) algebraic structure
which is generated by the operators
\begin{equation*}
\hat K_+=\sum_ma^\dag_{m+}a_{m-},
\qquad
\hat K_-=\left(\hat K_+\right)^\dag,
\qquad
\hat K_z={\frac12}(\hat n_+-\hat n_-),
\end{equation*}
written in terms of the creation and annihilation operators
$a^\dag_{m\sigma}$ and $a_{m\sigma}$,
with $m=1,\dots,\Omega$ and $\sigma=\pm$,
and where $\hat n_\pm$ counts the number of nucleons in the level with $\sigma=\pm$.
The hamiltonian
\begin{equation*}
\hat H=
\epsilon\hat K_z
+{\frac12}\upsilon\left(\hat K_+\hat K_-+\hat K_-\hat K_+\right)
+{\frac12}\omega\left(\hat K_+^2+\hat K_-^2\right),
\end{equation*}
can, with use of the underlying SU(2) algebra,
be solved analytically for certain values
of the parameters $\epsilon$, $\upsilon$ and $\omega$.
These have a simple physical meaning:
$\epsilon$ is the energy needed to promote
a nucleon from the lower level with $\sigma=-$
to the upper level with $\sigma=+$,
$\upsilon$ is the strength of the interaction that mixes configurations
with the same nucleon numbers $n_-$ and $n_+$,
and $\omega$ is the strength of the interaction
that mixes configurations differing by two in these numbers.
The Lipkin model has thus three ingredients
(albeit in schematic form)
that are of importance in determining the structure of nuclei:
an interaction $\upsilon$
between the nucleons in a valence shell,
the possibility to excite nucleons
from the valence shell into a higher shell
at the cost of an energy $\epsilon$,
and an interaction $\omega$
that mixes these particle--hole excitations
with the valence configurations.
With these ingredients
the Lipkin model has played an important role
as a testing ground of various approximations
proposed in nuclear physics,
examples of which are given in reference~\cite{Ring80}.

\section{Geometric collective models}
\label{s_gcm}
In 1879, in a study of the properties of a droplet of incompressible liquid,
Lord Rayleigh showed~\cite{Rayleigh79} that its normal modes of vibration
are described by the variables $\alpha_{\lambda\mu}$
which appear in the expansion of the droplet's radius,
\begin{equation}
R(\theta,\phi)=R_0\left(1+\sum_{\lambda\mu}\alpha_{\lambda\mu}^*Y_{\lambda\mu}(\theta,\phi)\right),
\label{e_multipole}
\end{equation}
where $Y_{\lambda\mu}(\theta,\phi)$ are {\sl spherical harmonics}
in terms of the {\sl spherical angles} $\theta$ and $\phi$.
Since the atomic nucleus from early on
was modeled as a dense, charged liquid drop~\cite{Weizsaecker35},
it was natural for nuclear physicists to adopt the same multipole parameterization~(\ref{e_multipole}),
as was done in the classical papers on the geometric collective model
by Rainwater~\cite{Rainwater50}, Bohr~\cite{Bohr52}, and Bohr and Mottelson~\cite{Bohr53}.

As was also shown by Lord Rayleigh,
the multipolarity that corresponds to the normal mode with lowest eigenfrequency
is of {\bf quadrupole} nature, $\lambda=2$.
The quadrupole collective coordinates $\alpha_{2\mu}$
can be transformed to an intrinsic-axes system
through $a_{2\mu}=\sum_\nu {\cal D}^2_{\nu\mu}(\theta_i)\alpha_{2\nu}$,
with ${\cal D}^2_{\nu\mu}(\theta_i)$ the {\sl Wigner $\cal D$ functions}
in terms of the {\sl Euler angles} $\theta_i$ that rotate the laboratory frame into the intrinsic frame.
If the intrinsic frame is chosen to coincide with the principal axes
of the quadrupole-deformed ellipsoid,
the $a_{2\mu}$ satisfy $a_{2-1}=a_{2+1}=0$ and $a_{2-2}=a_{2+2}$
while the remaining two variables
can be transformed further to two coordinates $\beta$ and $\gamma$,
according to $a_0=\beta\cos\gamma$ and $a_{2-2}=a_{2+2}=\beta\sin\gamma/\sqrt{2}$.
The coordinate $\beta\geq0$ parameterizes deviations from sphericity
while $\gamma$ is a polar coordinate confined to the interval $[0,\pi/3]$.
For $\gamma=0$ the intrinsic shape is axially symmetric and prolate,
for $\gamma=\pi/3$ it is axially symmetric and oblate,
and intermediate values of $\gamma$ describe triaxial shapes.

The classical problem of quadrupole oscillations of a droplet
has been quantized by Bohr~\cite{Bohr52},
resulting in the hamiltonian
\begin{equation*}
\hat H_{\rm B}=\hat T_\beta+\hat T_\gamma+\hat T_{\rm rot}+V(\beta,\gamma),
\end{equation*}
where $\hat T$ ($V$) refers to kinetic (potential) energy.
The kinetic energy has three contributions
coming from $\beta$ oscillations which preserve axial symmetry,
from $\gamma$ oscillations which do not
and from the rotation of a quadrupole-deformed object.
Bohr's analysis results in a collective Schr\"odinger equation
$\hat H_{\rm B}\Psi(\beta,\gamma,\theta_i)=E\Psi(\beta,\gamma,\theta_i)$ with
\begin{eqnarray}
\hat H_{\rm B}&=&-\frac{\hbar^2}{2B_2}
\left[\frac{1}{\beta^4}\frac{\partial}{\partial\beta}\beta^4\frac{\partial}{\partial\beta}+
\frac{1}{\beta^2\sin3\gamma}\frac{\partial}{\partial\gamma}\sin3\gamma\frac{\partial}{\partial\gamma}\right]
\nonumber\\&&
+\frac{\hbar^2}{8B_2}\sum_{k=1}^3\frac{\hat L_k^{\prime2}}{\beta^2\sin^2(\gamma-2\pi k/3)},
\label{e_bohr}
\end{eqnarray}
where $B_2=\rho R_0^5/2$ is the mass parameter
in terms of the constant matter density $\rho$ for an incompressible nucleus.
The operators $\hat L_k^\prime$ are the components of the angular momentum
in the intrinsic frame of reference
where the prime is used to distinguish these
from the components of the angular momentum
in the laboratory frame of reference.
The collective coordinates
are coupled in an intricate way in the Bohr hamiltonian~(\ref{e_bohr})
and this limits the number of exactly solvable cases.
In particular, because of the $\gamma$ dependence of the moments of inertia,
$\gamma$ excitations are strongly coupled to the collective rotational motion.
It turns out that $\beta$ excitations are less strongly coupled
and a judicious choice of the potential may well lead to a separation 
of $\beta$ from the $\gamma$ and $\theta_i$ coordinates.

\subsection{Exactly solvable collective models}
\label{s_exact}
A way to decouple the Bohr hamiltonian~(\ref{e_bohr})
into separate differential equations
was proposed by Wilets and Jean~\cite{Wilets56}
and requires a potential of the form 
\begin{equation*}
V(\beta,\gamma)=V_1(\beta)+\frac{V_2(\gamma)}{\beta^2},
\end{equation*}
leading to the coupled equations
\begin{eqnarray}
&\Bigg[-\frac{1}{\beta^4}\frac{\partial}{\partial\beta}\beta^4\frac{\partial}{\partial\beta}
+u_1(\beta)-\varepsilon +\frac{\omega}{\beta^2}\Bigg]\xi(\beta)=0,
\label{e_beta}\\
&\Bigg[-\frac{1}{\sin3\gamma}\frac{\partial}{\partial\gamma}\sin3\gamma\frac{\partial}{\partial\gamma}+
\sum_{k=1}^3\frac{\hat L^{\prime2}_k}{4\sin^2(\gamma-2\pi k/3)}
+u_2(\gamma)-\omega\Bigg]\psi(\gamma,\theta_i)=0,
\nonumber\\&&
\label{e_gamma}
\end{eqnarray}
where $\omega$ is the separation constant,
$\varepsilon=(2B_2/\hbar^2)E$ and $u_i=(2B_2/\hbar^2)V_i$ ($i=1,2$).
The first equation can only be solved exactly
if the constant $\omega$ is obtained
from the solution of the second one. 
At present the only known analytic solution of the Bohr hamiltonian~(\ref{e_bohr})
is for {\bf $\gamma$-independent} potentials~\cite{Wilets56},
that is, for $V_2(\gamma)=0$.
In that case, one still needs 
to determine the allowed values of $\omega$ in the equation~(\ref{e_gamma}).
Many techniques have been proposed to solve this equation
relying on either algebraic or analytic methods.
Rakavy~\cite{Rakavy57} noticed that the first two terms in equation~(\ref{e_gamma})
correspond to the Casimir operator of the orthogonal group in five dimensions, SO(5),
and it is known from group-theoretical arguments
that $\omega$ therefore acquires the values $\omega=v(v+3)$ with $v=1,2,\dots$,
leading to the following equation in $\beta$:
\begin{equation}
\left[-\frac{1}{\beta^4}\frac{\partial}{\partial\beta}\beta^4\frac{\partial}{\partial\beta}
+u_1(\beta)-\varepsilon +\frac{v(v+3)}{\beta^2}\right]\xi(\beta)=0.
\label{e_beta2}
\end{equation}
Special choices of $u_1(\beta)$ [or $V_1(\beta)$] lead to
the following exact solutions of the Bohr hamiltonian~(\ref{e_bohr}).
 
\subsubsection{The five-dimensional harmonic oscillator.}
\label{sss_ho}
The {\bf harmonic quadrupole oscillator} was the first potential used
in an exactly solvable collective model~\cite{Bohr52}. 
The potential $V(\beta,\gamma)$ reduces to a single term $V(\beta)=C_2\beta^2/2$
where $C_2$ is a constant.
Even though one does not expect harmonic quadrupole vibrations
to appear in the experimental study of atomic nuclei,
the model serves as an interesting benchmark. 
The solution of the differential equation in $\beta$
results in the energy spectrum $E(n,v)=\hbar\Omega(2n+v+5/2)$
with $\Omega=\sqrt{C_2/B_2}$
and the corresponding eigenfunctions
are {\sl associated Legendre polynomials} of order $v+3/2$.
%(see Arfken and Weber~\cite{Arf01}).
The energy spectrum is characterized by degeneracies
that increase with increasing $n$ and $v$.
The complete solution of the Bohr hamiltonian with a harmonic potential
can be obtained with group-theoretical methods
based on the reduction ${\rm U}(5)\supset{\rm SO}(5)\supset{\rm SO}(3)$~\cite{Chacon76}.
An alternative derivation is based on the notion of quasi-spin
discussed in subsection~\ref{ss_racah0} 
which for bosons has the algebraic structure is SU(1,1)~\cite{Rowe98}.

\subsubsection{The infinite square-well potential.}
\label{sss_e5}
It was shown by Wilets and Jean ~\cite{Wilets56} that
the spectrum of the five-dimensional harmonic oscillator
can be made anharmonic by introducing a potential in $\beta$
that has the form of an {\bf infinite square well}, that is,
$V(\beta)={\rm constant}$  for $\beta\leq b$
and $V(\beta)=\infty$ for $\beta>b$.
This leads to solutions of equation~(\ref{e_beta2})
that are Bessel functions with allowed values for $v$ 
resulting from the boundary condition of a vanishing wave function at $\beta=b$. 

The solution of this problem has been worked out much later by Iachello~\cite{Iachello00}
in the context of a study on shape transitions
from spherical and to $\gamma$-soft potentials.
The spectrum is determined by the energy eigenvalues
\begin{equation*}
E(i,v)=\frac{\hbar^2}{2B_2}k^2_{i,v},
\qquad
k_{i,v}=\frac{x_{i,v}}{b},
\end{equation*}
with corresponding eigenfunctions
\begin{equation*}
\xi_{i,v}(\beta)\propto
\beta^{-3/2}J_{v+3/2}(k_{i,v}\beta),
\end{equation*} 
where $x_{i,v}$ is the $i$th zero of the Bessel function $J_{v+3/2}(x)$.
This solution, referred to as {\bf E(5)}, proves therefore to be exact,
as discussed in great detail in reference~\cite{Iachello00}. 
 
\subsubsection{The Davidson potential.}
\label{sss_davidson}
\begin{figure}
\centering
\includegraphics[width=8cm]{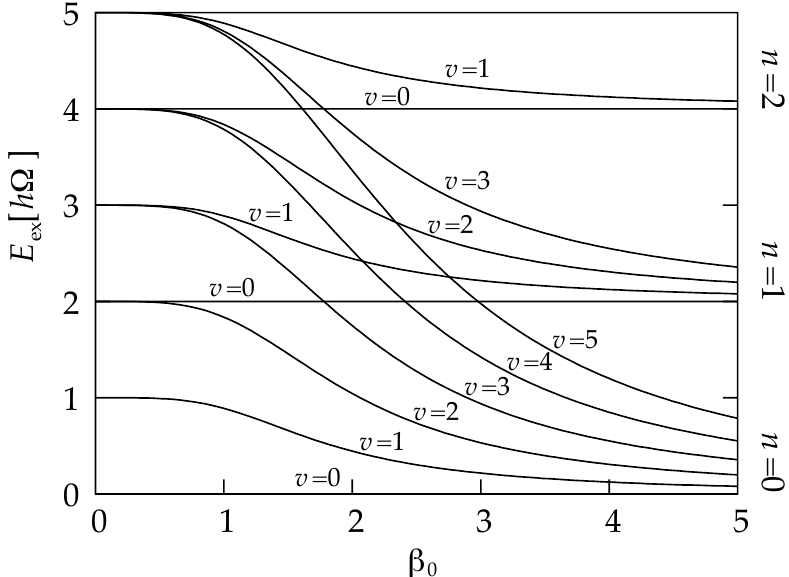}
\caption{The energy spectrum $E(n,\tilde v)=2n+\tilde v+5/2$
(in units $\hbar\Omega$) of the Davidson potential 
as a function of the deformation parameter $\beta_0$.
Taken from reference~\cite{Baerdemacker07a}.}
\label{f_davidson}
\end{figure}

The five-dimensional analogue of a three-dimensional potential,
proposed by {\bf Davidson}~\cite{Davidson32} for use in molecular physics,
gives rise to another analytic solution of the Bohr hamiltonian.
The constraint of $\gamma$ independence is kept
and the harmonic potential of subsection~\ref{sss_ho} is modified to
$V(\beta)=C_2(\beta^2+\beta_0^4/\beta^2)/2$.
The additional term changes the spherical potential
into a deformed one with a minimum value located at $\beta_0$.
The energy spectrum of the modified potential
can be obtained from the spherical one
after the substitution $v\mapsto\tilde v$,
with $\tilde v$ defined from $\tilde v(\tilde v+3)=v(v+3)+k\beta_0^4$
with $k=B_2C_2/\hbar^2$.
The resulting energy spectrum is shown in figure~\ref{f_davidson}.
The corresponding problem with a mass parameter $B_2$
depending on the coordinate $\beta$ has also been studied~\cite{Bonatsos10}.
If one considers the form $B_2=B_2(0)/(1+a\beta^2)^2$,
the problem becomes exactly solvable
with use of techniques from supersymmetric quantum mechanics~\cite{Cooper95}
and by imposing integrability conditions, also called shape invariance~\cite{Bagchi05}.

\subsubsection{Other analytic solutions.}
\label{sss_other}
There are other $\gamma$-independent potentials $V(\beta)$
that lead to a solvable equation~(\ref{e_beta2})
and therefore yield an exactly solvable Bohr hamiltonian.
Most notably, they are the {\bf Coulomb} potential $V(\beta)=-A/\beta$
and the {\bf Kratzer} potential $V(\beta)=-B[\beta_0/\beta-\beta^2_0/(2\beta^2)]$~\cite{Fortunato03}.
Also potentials of the form $V(\beta)=\beta^{2n}$ ($n=1,2,\dots$)
have been studied,
which for $n=1$ reduce to the five-dimensional harmonic oscillator
and for $n\rightarrow\infty$ approach the infinite square-well potential,
but with numerical techniques
(see, {\it e.g.}, the reviews~\cite{Fortunato05,Bonatsos07}).
L\'evai and Arias~\cite{Levai04} proposed a sextic potential
leading to a quasi-exactly solvable model~\cite{Turbiner88,Ushveridze94}
which reduces to a class of two-parameter potentials
containing terms in $\beta^2$, $\beta^4$ and $\beta^6$. 
This choice leads to exact solutions of the Bohr hamiltonian for a finite subset of states,
here in particular for the lowest few eigenstates
(energies, wave functions and a subset of $B$(E2) values).
Finally, the particular choice $V(\beta)=C_2b\beta^2/(1+b\beta^2)$,
proposed by Ginocchio~\cite{Ginocchio80}, is solvable.
It leads to a solution of the Bohr hamiltonian
that reproduces the lowest energy eigenvalues of an anharmonic vibrator
[or of the U(5) limit of the IBM, see section~\ref{s_ibm}].

\subsection{Triaxial models}
\label{ss_triaxial}
Many nuclei may exhibit excursions away from axial symmetry,
requiring the introduction of explicit {\bf triaxial} features in the Bohr hamiltonian.
Due to the coupling of vibrational and rotational degrees of freedom in the Bohr hamiltonian,
potentials with $\gamma$ dependence allow very few exact solutions,
even if they are of the separable type, $V(\beta,\gamma)=V_1(\beta)+V_2(\gamma)$.
In early attempts to address this more complicated situation,
triaxial rotors were studied in an adiabatic approximation
which implies that the nucleus' intrinsic shape does not change under the effect of rotation.
Such systems, in the context of the Bohr hamiltonian,
correspond to a potential of the type 
$V(\beta,\gamma)=\delta(\beta-\beta_0)\delta(\gamma-\gamma_0)$
and their hamiltonian contains a rotational kinetic energy term only.  
On the other hand, the quantum mechanics of a rotating rigid body
was studied much before the advent of the Bohr hamiltonian,
by Reiche~\cite{Reiche26} and by Casimir~\cite{Casimir31},
starting from a classical description of rotating bodies.
The two approaches give rise to rather different moments of inertia,
as discussed in the next subsection~\ref{sss_rotor}.

\subsubsection{Rigid rotor models.}
\label{sss_rotor}
{\bf Davydov} and co-workers~\cite{Davydov58,Davydov59}
studied and solved a triaxial rotor model in the context of the Bohr hamiltonian,
which in the adiabatic approximation reduces to its rotational part,
\begin{equation}
\hat H_{\rm rot}=
\frac{\hbar^2}{8B_2}\sum_{k=1}^3\frac{\hat L_k^{\prime2}}{\beta_0^2\sin^2(\gamma_0-2\pi k/3)},
\label{e_davydov}
\end{equation}
where $\beta_0$ and $\gamma_0$ are {\em fixed} values
that define the shape of the rotating nucleus.
The dependence of the moments of inertia 
${\cal J}_k=4B_2\beta_0^2\sin^2(\gamma_0-2\pi k/3)$
on the shape parameters $\beta_0$ and $\gamma_0$
is that of a droplet in {\bf irrotational flow},
that is, of which the velocity field $\bar v(\bar r)$
obeys the condition $\bar\nabla\wedge\bar v(\bar r)=0$.

The Davydov model is exactly solvable in the sense
that the energies of the lowest-spin states $L^\pi=0^+,2^+,3^+,\dots$
can be derived in closed form.
For higher-spin states the energies
are obtained as solutions of higher-order algebraic equations:
cubic for $L^\pi=4^+$, quartic for $L^\pi=6^+$, {\it etc.}
The corresponding wave functions only depend on the Euler angles $\theta_i$
and can be expressed as $\Phi_{iLM}(\theta_i)=\sum_Ka_K^i\Phi_{KLM}(\theta_i)$,
with coefficients $a_K^i$ obtained from the same algebraic equations, and
\begin{equation*}
\Phi_{KLM}(\theta_i)=
\sqrt{\frac{2L+1}{16\pi^2(1+\delta_{K0})}}
\left[{\cal D}_{MK}^L(\theta_i)+(-)^L{\cal D}_{M,-K}^L(\theta_i)\right],
\end{equation*}
where ${\cal D}_{MK}^L(\theta_i)$ are the {\sl Wigner $\cal D$ functions}.
These expressions also allow the calculation of electromagnetic transitions~\cite{Davydov58}.

\begin{figure}
\centering
\includegraphics[angle=0,width=12cm]{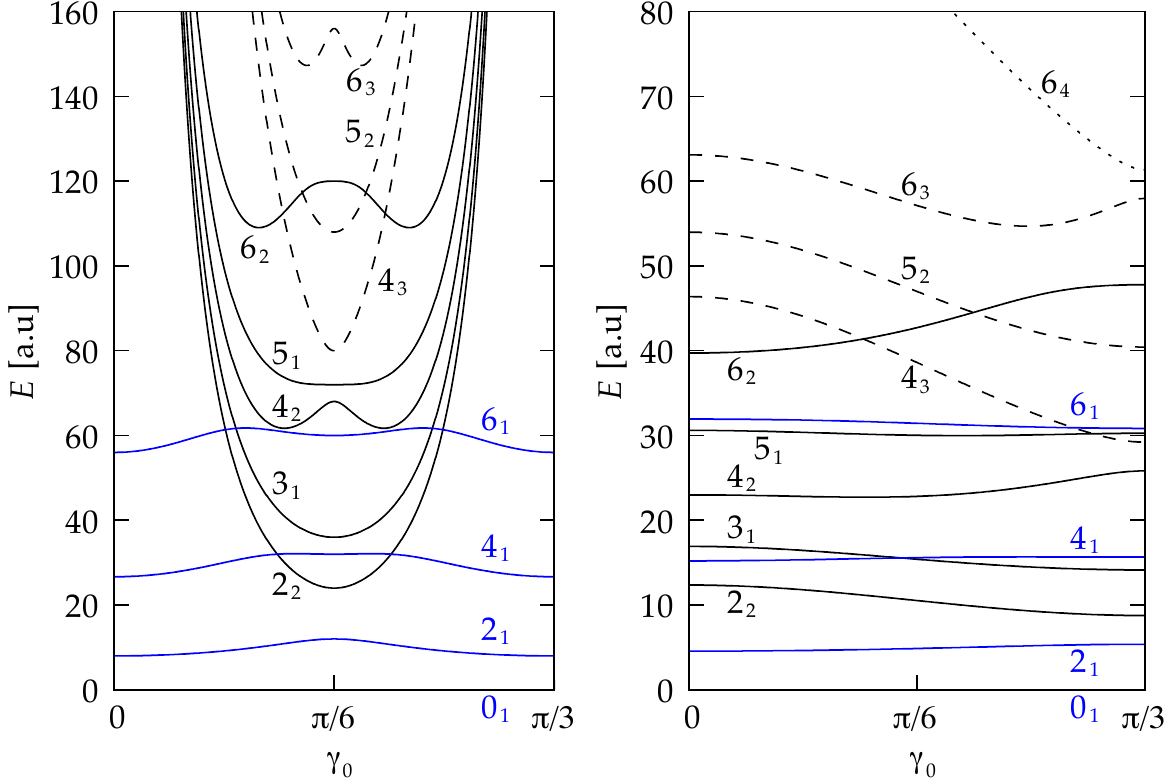}
\caption{
The energy spectrum of a rigid rotor,
with irrotational (left) and rigid (right) values for the moments of inertia
and with energy in units $\hbar^2/(8B_2)$ and $5\hbar^2/(4m_{\rm n}AR_0^2)$, respectively.
In both cases $\beta_0=1$.
Taken from reference~\cite{Baerdemacker07a}.}
\label{f_davydov}
\end{figure}
The classical expressions for the moments of inertia
of a {\bf rigid body} with quadrupole deformation, on the other hand, are 
${\cal J}_k=(2m_{\rm n}AR_0^2/5)[1-\sqrt{5/4\pi}\beta_0\cos(\gamma_0-2\pi k/3)]$
where $m_{\rm n}A$ is the mass.
As a result, its quantum-mechanical rotation
leads to an energy spectrum~\cite{Reiche26,Casimir31}
which is different from the one obtained with the Bohr hamiltonian (see figure~\ref{f_davydov}).
The most obvious difference between the two cases
occurs in the limits of axial symmetry ($\gamma_0=0$ or $\gamma_0=\pi/3$)
when one of the moments of inertia diverges in the Davydov model.
This divergence results from the extreme picture of rigid rotation
and disappears when the rigid triaxial rotor model is generalized
by allowing softness in the $\beta$ and $\gamma$ degrees of freedom~\cite{Davydov60,Davydov61}.

\subsubsection{The Meyer-ter-Vehn model.}
\label{sss_meyer}
{\bf Meyer-ter-Vehn found} an interesting solution of a rigid rotor with $\gamma_0=\pi/6$~\cite{Meyer75}.
For this value of $\gamma_0$ the moments of inertia  ${\cal J}_2$ and ${\cal J}_3$
in equation~(\ref{e_davydov}) are equal
while the three intrinsic quadrupole moments are different.
The hamiltonian~(\ref{e_davydov}) can then be rewritten
in the form $\hbar^2/(2B_2\beta_0^2)(\hat L^{\prime2}-3\hat L_1^{\prime2}/4)$
with energy eigenvalues $\hbar^2/(2B_2\beta_0^2)[L(L+1)-3R^2/4]$,
where $L$ denotes the angular momentum
and $R$ the projection of $L$ on the 1-axis
(perpendicular to the 3-axis) which is a good quantum number for such systems.
This model can be used for odd-mass nuclei
by coupling an odd particle to the triaxial rotor~\cite{Meyer75}.

\subsubsection{Approximate solutions for soft potentials.}
\label{sss_asoft}
While the rigid rotor may serve
as a good starting point for the description of certain nuclei,
the strong coupling between $\gamma$ excitations and the collective rotational motion
calls for simple, more realistic models,
in particular for strongly deformed nuclei in the rare-earth and actinide regions.
One approach is to assume harmonic-oscillator (or other schematic) potentials 
in the $\gamma$ and $\beta$ variables,
such that the Bohr hamiltonian can be solved approximately.
Even with potentials of the Wilets--Jean type
that allow an exact decoupling of the $\beta$ degree of freedom,
an analytic solution of the $(\gamma,\theta_i)$ part of the wave function
requires moments of inertia frozen at a certain $\gamma_0$ value
[corresponding with the minimum of the $V(\gamma)$ potential]
in addition to the assumption of harmonic motion around $\gamma_0$. 
With these restrictions analytic solutions can be obtained.
Bonatsos {\it et.al.} studied a large number of such potentials,
deriving special solutions of the Bohr Hamiltonian
characterized by various expressions of $V_1(\beta)$ and $V_2(\gamma)$
(see the review paper~\cite{Bonatsos07}).
The validity of these approximations has to be confronted
with numerical studies (see section~\ref{ss_algebraic}). 
Two particular approximate analytic solutions,
extensively confronted with experimental data in the rare-earth region,
are named {\bf X(5)}~\cite{Iachello01} and {\bf Y(5)}~\cite{Iachello03}.
The corresponding potentials  
which are separable in $\beta$ and $\gamma$,
make use of a square-well potential in the $\beta$ direction
and a harmonic oscillator in the  $\gamma$ direction $\propto\gamma^2$ [for the X(5) solution]
and of a harmonic oscillator in the $\beta$ direction $\propto(\beta-\beta_0)^2$ 
and an infinite square-well potential in the $\gamma$ direction
around $\gamma=0$ [for the Y(5) solution].

Many other models exhibiting softness in both the $\beta$ and $\gamma$ degrees of freedom
are discussed in the review papers by Fortunato~\cite{Fortunato05} and Cejnar {\it et al.}~\cite{Cejnar10}.

\subsubsection{Partial solutions.}
\label{sss_partial}
There are some models that can be solved exactly for a limited number of states.
An example is the {\bf P\"oschl--Teller} potential
$V(\gamma)=a/\sin^23\gamma$
which has an exact solution for the $J=0$ and $J=3$ states~\cite{Baerdemacker06}.

\subsection{Geometric collective models: an algebraic approach}
\label{ss_algebraic}
Exactly solvable models are only possible for specific potentials $V(\beta,\gamma)$
and are clearly limited in scope.
To handle a general potential $V(\beta,\gamma)$,
the differential equation associated with the Bohr hamiltonian~(\ref{e_bohr})
must be solved numerically~\cite{Gneuss71}.

An algebraic approach based on ${\rm SU}(1,1)\otimes{\rm SO}(5)$,
has been proposed by Rowe~\cite{Rowe04a}
and, independently, by De Baerdemacker {\it et al.}~\cite{Baerdemacker07}.
To improve the convergence in a five-dimensional oscillator basis,
a direct product is taken
of SU(1,1) wave functions in $\beta$
with ${\rm SO}(5)\supset{\rm SO}(3)$ generalized spherical harmonics
in $\gamma$ and $\theta_i$.
This algebraic structure allows the calculation
of a general set of matrix elements
of potential and kinetic energy terms in closed analytic form.
Consequently, the exact solutions of the harmonic oscillator,
the $\gamma$-independent rotor
and the axially deformed rotor can be derived easily.
As a nice illustration of this approach,
the solution of the Davidson potential (see section~\ref{sss_davidson})
can be obtained in the closed form~\cite{Rowe98}.
The strength of this approach (also called the algebraic collective model) 
is that one can go beyond the adiabatic separation 
of the $\beta$ and $\gamma$ vibrational modes,
usually taken as harmonic, and test this restriction
(see, {\it e.g.}, reference~\cite{Caprio11}).
Presently, more realistic potential and kinetic energy terms are considered,
leading to numerical studies going far
beyond the constraints of the exactly solvable models considered here. 

\section{The interacting boson model}
\label{s_ibm}
In the geometric collective model exact solutions
are found for specific potentials in the Bohr hamiltonian~(\ref{e_bohr}).
They correspond to solutions of coupled differential equations
in terms of standard mathematical functions
and have no obvious connection with the algebraic formulation
of the quantal $n$-body problem of section~\ref{s_nbody}.
Alternatively, collective nuclear excitations can be described
with the interacting boson model ({\bf IBM}) of Arima and Iachello~\cite{Arima75}
which, in contrast, can be formulated in an algebraic language.

The original version of the IBM, applicable to even--even nuclei,
describes nuclear properties in terms of interacting $s$ and $d$ bosons
with angular momentum $\ell=0$ and $\ell=2$,
and a vacuum state $|{\rm o}\rangle$
which represents a doubly-magic core.
Unitary transformations among the six states
$s^\dag|{\rm o}\rangle$ and $d^\dag_m|{\rm o}\rangle,m=0,\pm1,\pm2$,
also collectively denoted by $b_{\ell m}^\dag$,
generate the Lie algebra U(6) (see section~\ref{s_nbody}).

In nuclei with many valence neutrons and protons,
the dimension of the shell-model space is prohibitively large.
A drastic reduction of this dimension is obtained
if shell-model states are considered
that are constructed out of nucleon pairs
coupled to angular momenta $J=0$ and $J=2$ only.
If, furthermore, a mapping is carried out
from nucleon pairs to genuine $s$ and $d$ bosons,
a connection between the shell model and the IBM is established~\cite{Otsuka78}.

Given this microscopic interpretation of the bosons,
a low-lying collective state of an even--even nucleus
with $2N_{\rm b}$ valence nucleons
is approximated as an $N_{\rm b}$-boson state.
Although the separate boson numbers $n_s$ and $n_d$
are not necessarily conserved,
their sum $n_s+n_d=N_{\rm b}$ is.
This implies a hamiltonian that conserves the total boson number,
of the form $\hat H_{\rm IBM}=E_0+\hat H_1+\hat H_2+\hat H_3+\cdots$,
where the index refers to the order of the interaction
in the generators of U(6)
and where the first term is a constant
which represents the binding energy of the core. 

The characteristics of the most general IBM hamiltonian
which includes up to two-body interactions
and its group-theoretical properties are well understood~\cite{Castanos79}.
Numerical procedures exist to obtain its eigensolutions
but, as in the nuclear shell model,
this quantum-mechanical many-body problem
can be solved analytically for particular choices
of boson energies and boson--boson interactions.
For an IBM hamiltonian with up to two-body interactions between the bosons,
three different analytical solutions or limits exist:
the {\bf vibrational} U(5)~\cite{Arima76},
the {\bf rotational} SU(3)~\cite{Arima78}
and the {\bf $\gamma$-unstable} SO(6) limit~\cite{Arima79}.
They are associated with the following lattice of algebras:
\begin{equation}
{\rm U}(6)\supset
\left\{\begin{array}{c}
{\rm U}(5)\supset{\rm SO}(5)\\
{\rm SU}(3)\\
{\rm SO}(6)\supset{\rm SO}(5)
\end{array}\right\}
\supset{\rm SO}(3).
\label{e_ibmlat}
\end{equation}
The algebras appearing in the lattice~(\ref{e_ibmlat}) are subalgebras of U(6)
generated by operators of the type $b^\dag_{\ell m}b_{\ell'm'}$.
If the energies and interactions are chosen such
that $\hat H_{\rm IBM}$ reduces to a sum of Casimir operators of subalgebras
belonging to a chain of {\em nested} algebras in the lattice~(\ref{e_ibmlat}),
the eigenvalue problem,
according to the discussion of section~\ref{s_nbody},
can be solved analytically
and the quantum numbers
associated with the different Casimir operators are conserved.

An important aspect of the IBM is its geometric interpretation
which can be obtained
by means of {\bf coherent} (or intrinsic) states~\cite{Ginocchio80b,Dieperink80,Bohr80}.
The ones used for the IBM are of the form
\begin{equation*}
|N;\alpha_{2\mu}\rangle\propto
\left(s^\dag+\sum_\mu\alpha_{2\mu}d^\dag_\mu\right)^N|{\rm o}\rangle,
\end{equation*}
where the $\alpha_{2\mu}$ are similar to the shape variables
of the geometric collective model (see section~\ref{s_gcm}).
In the same way as in that model,
the $\alpha_{2\mu}$ can be related to Euler angles $\theta_i$
and two intrinsic shape variables, $\beta$ and $\gamma$,
that parameterize quadrupole vibrations
of the nuclear surface around an equilibrium shape.
The expectation value of an operator in the coherent state
leads to a functional expression in $N$, $\beta$ and $\gamma$.
The most general IBM hamiltonian, therefore,
can be converted in a total energy surface $E(\beta,\gamma)$.
An analysis of this type shows that
the three limits of the IBM
have simple geometric counterparts
that are frequently encountered in nuclei~\cite{Ginocchio80b,Dieperink80}.

\subsection{Neutrons and protons: $F$ spin}
\label{ss_fspin}
The recognition that the $s$ and $d$ bosons can be identified
with pairs of valence nucleons coupled to angular momenta $J=0$ or $J=2$,
made it clear that a connection between the boson and shell model
required a distinction between neutrons and protons.
Consequently, an extended version of the model
was proposed by Arima {\it et al.}~\cite{Arima77}
in which this distinction was made, referred to as \mbox{\bf IBM-2},
as opposed to the original version of the model, \mbox{\bf IBM-1}.

In the \mbox{IBM-2} the total number of bosons $N_{\rm b}$
is the sum of the neutron and proton boson numbers,
$N_\nu$ and $N_\pi$, which are conserved separately.
The algebraic structure of \mbox{IBM-2}
is a product of U(6) algebras, ${\rm U}_\nu(6)\otimes{\rm U}_\pi(6)$,
consisting of the operators $b^\dag_{\nu,\ell m}b_{\nu,\ell'm'}$ for the neutron bosons
and $b^\dag_{\pi,\ell m}b_{\pi,\ell'm'}$ for the proton bosons.
The model space of \mbox{IBM-2}
is the product of symmetric irreducible representations
$[N_\nu]\times[N_\pi]$ of ${\rm U}_\nu(6)\otimes{\rm U}_\pi(6)$.
In this model space the most general, $(N_\nu,N_\pi)$-conserving,
rotationally invariant \mbox{IBM-2} hamiltonian is diagonalized.

The \mbox{IBM-2} proposes a phenomenological description
of low-energy collective properties of medium-mass and heavy nuclei.
In particular, energy spectra and E2 and M1 transition properties
can be reproduced with a global parameterization as a function
of the number of valence neutrons and protons
but the detailed description of specific nuclear properties can remain a challenge.
The classification and analysis of the symmetry limits of \mbox{IBM-2}
is considerably more complex
than the corresponding problem in \mbox{IBM-1}
but are known for the most important limits
which are of relevance in the analysis of nuclei~\cite{Isacker86}.

The existence of two kinds of bosons offers the possibility to assign
an {\bf $F$-spin} quantum number to them, $F={\frac12}$,
the boson being in two possible charge states
with $M_F=-{\frac12}$ for neutrons
and $M_F=+{\frac12}$ for protons~\cite{Otsuka78}.
Formally, $F$ spin is defined by the algebraic reduction
\begin{equation*}
\begin{array}{ccccccc}
{\rm U}(12)&\supset&{\rm U}(6)&\otimes&\Big({\rm U}(2)&\supset&{\rm SU}(2)\Big)\\
\downarrow&&\downarrow&&\downarrow&&\downarrow\\[0mm]
[N_{\rm b}]&&[N_{\rm b}-f,f]&&[N_{\rm b}-f,f]&&F
\end{array},
\end{equation*}
with $2F$ being the difference between the labels
that characterize U(6) or U(2),
$F=[(N_{\rm b}-f)-f]/2=(N_{\rm b}-2f)/2$.
The algebra U(12) consists of the generators
$b^\dag_{\rho,\ell m}b_{\rho',\ell'm'}$, with $\rho,\rho'=\nu$ or $\pi$,
which also includes operators
that change a neutron boson into a proton boson or {\it vice versa} ($\rho\neq\rho'$).
Under this algebra U(12) bosons behave symmetrically
whence the symmetric irreducible representation $[N_{\rm b}]$.
The irreducible representations of U(6) and U(2), in contrast, do not have to be symmetric
but, to preserve the overall U(12) symmetry, they should be identical.

The mathematical structure of $F$ spin
is entirely similar to that of isospin $T$.
An $F$-spin SU(2) algebra can be defined
which consists of the diagonal operator $\hat F_z=(-\hat N_\nu+\hat N_\pi)/2$
and the raising and lowering operators $\hat F_\pm$
that transform neutron into proton bosons or {\it vice versa}.
These are the direct analogues
of the isopin generators $\hat T_z$ and $\hat T_\pm$.
The physical meaning of $F$ spin and isospin is different, however,
as the mapping of a shell-model hamiltonian with isospin symmetry
does not necessarily yield an $F$-spin conserving hamiltonian in \mbox{IBM-2}.
Conversely, an $F$-spin conserving \mbox{IBM-2} hamiltonian
may or may not have eigenstates with good isospin.
If the neutrons and protons occupy different shells,
so that the bosons are defined in different shells,
then {\em any} \mbox{IBM-2} hamiltonian has eigenstates
that correspond to shell-model states with good isospin,
irrespective of its $F$-spin symmetry character.
If, on the other hand, neutrons and protons occupy the same shell,
a general \mbox{IBM-2} hamiltonian does {\em not} lead to states with good isospin.
The isospin symmetry violation is particularly significant in nuclei
with approximately equal numbers of neutrons and protons ($N\sim Z$)
and requires the consideration of \mbox{IBM-3} (see section~\ref{ss_isospin}).
As the difference between the numbers of neutrons and protons in the same shell increases,
an approximate equivalence of $F$ spin and isospin is recovered
and the need for \mbox{IBM-3} disappears~\cite{Elliott87}.

Just as {\sl isobaric multiplets} of nuclei are defined through the connection
implied by the raising and lowering operators $\hat T_\pm$,
{\bf $F$-spin multiplets} can be defined through the action of $\hat F_\pm$~\cite{Brentano85}.
The states connected are in nuclei with $N_\nu+N_\pi$ constant;
these can be isobaric (constant nuclear mass number $A$)
or may differ by multiples of $\alpha$ particles,
depending on whether the neutron and proton bosons
are of the same or of a different type
(which refers to their particle- or hole-like character).

The phenomenology of $F$-spin multiplets is similar to that of isobaric multiplets
but for one important difference.
The nucleon--nucleon interaction favours spatially symmetric configurations
and consequently nuclear excitations at low energy
generally have $T=T_{\rm min}=|(N-Z)/2|$.
Boson--boson interactions also favour spatial symmetry
but that leads to low-lying levels with $F=F_{\rm max}=(N_\nu+N_\pi)/2$.
As a result, in the case of an $F$-spin multiplet
a relation is implied between the low-lying spectra of the nuclei in the multiplet,
while an isobaric multiplet (with $T\geq1$)
involves states at higher excitation energies in some nuclei.

Another important aspect of \mbox{IBM-2} is that it predicts states
that are additional to those found in \mbox{IBM-1}.
Their structure can be understood as follows.
States with maximal $F$ spin, $F=N/2$, are symmetric in U(6)
and are the exact analogues of \mbox{IBM-1} states.
The next class of states has $F=N/2-1$,
no longer symmetric in U(6) but belonging to its irreducible representation $[N-1,1]$.
Such states were studied theoretically in 1984 by Iachello~\cite{Iachello84}
and were observed, for the first time in $^{156}$Gd~\cite{Bohle84},
and later in many other deformed as well as spherical nuclei.

\begin{figure}
\centering
\includegraphics[width=10cm]{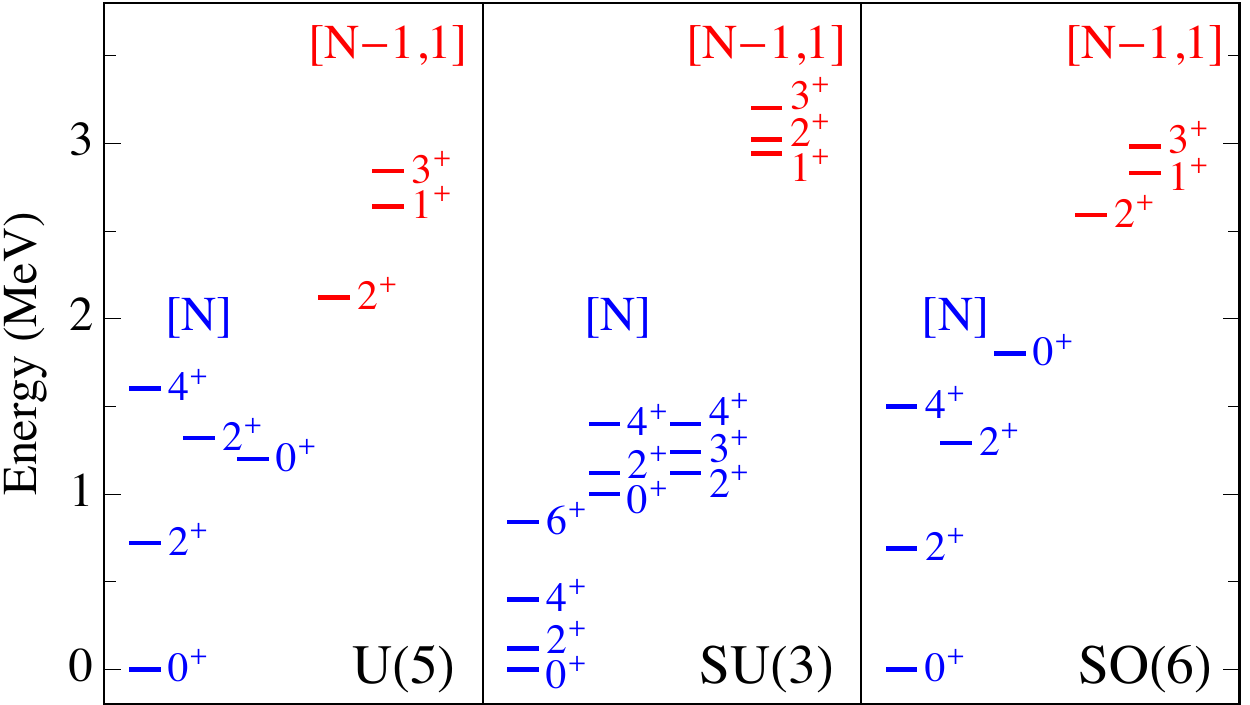}
\caption{
Partial energy spectra in the three limits of the \mbox{IBM-2}
in which $F$ spin is a conserved quantum number.
Levels are labelled by their angular momentum and parity $J^\pi$;
the U(6) labels $[N-f,f]$ are also indicated.
States symmetric in U(6) are in blue
while mixed-symmetry states are in red.}
\label{f_ms}
\end{figure}
The existence of these states with {\bf mixed symmetry},
excited in a variety of reactions,
is by now well established~\cite{Heyde10}.
The pattern of the lowest symmetric and mixed-symmetric states
is shown in figure~\ref{f_ms}.
Of particular relevance are $1^+$ states,
since these are allowed in \mbox{IBM-2}
but not in \mbox{IBM-1}.
The characteristic excitation of $1^+$ levels
is of magnetic dipole type
and the \mbox{IBM-2} prediction for the M1 strength
to the $1^+$ mixed-symmetry state is~\cite{Isacker86}
\begin{equation*}
B({\rm M}1;0^+_1\rightarrow1^+_{\rm MS})=
{3\over{4\pi}}(g_\nu-g_\pi)^2f(N)N_\nu N_\pi,
\end{equation*}
where $g_\nu$ and $g_\pi$ are the boson $g$ factors.
The function $f(N)$ is known analytically
in the three principal limits of the \mbox{IBM-2},
$f(N)=0$, $8/(2N-1)$ and $3/(N+1)$ in U(5), SU(3) and SO(6), respectively.
This gives a simple and reasonably accurate estimate
of the total M1 strength of orbital nature
to $1^+$ mixed-symmetry states in even--even nuclei.

The geometric interpretation of mixed-symmetry states can be found
by taking the limit of large boson number~\cite{Ginocchio92}.
From this analysis emerges
that they correspond to oscillations
in which the neutrons and protons are out of phase,
in contrast to the symmetric \mbox{IBM-2} states
for which such oscillations are in phase.
The occurrence of such states was first predicted
in the context of geometric two-fluid models
in vibrational~\cite{Faessler66} and deformed~\cite{Iudice78} nuclei
in which they appear
as neutron--proton counter oscillations.
Because of this geometric interpretation,
mixed-symmetry states are often referred to as {\bf scissors} states
which is the pictorial image one has
in the case of deformed nuclei.
The \mbox{IBM-2} thus confirms these geometric descriptions
but at the same time generalizes them to {\em all} nuclei,
not only spherical and deformed,
but $\gamma$ unstable and transitional as well.

\subsection{Neutrons and protons: Isospin}
\label{ss_isospin}
If neutrons and protons occupy different valence shells,
it is natural to consider neutron--neutron and proton--proton pairs only,
and to include the neutron--proton interaction explicitly between the two types of pairs.
If neutrons and protons occupy the same valence shell,
this approach no longer is valid
since there is no reason not to include the $T=1$ neutron--proton pair.
The ensuing model, proposed by Elliott and White~\cite{Elliott80}, is called \mbox{\bf IBM-3}.
Because the \mbox{IBM-3} includes the complete $T=1$ triplet,
it can be made isospin invariant,
enabling a more direct comparison with the shell model.

In the \mbox{IBM-3} there are three kinds of bosons
($\nu$, $\delta$ and $\pi$)
each with six components
and, as a result, an $N_{\rm b}$-boson state
belongs to the symmetric irreducible representation $[N_{\rm b}]$ of U(18).
It is possible to construct \mbox{IBM-3} states
that have good total angular momentum $J$
and good total isospin $T$.

The classification of dynamical symmetries of \mbox{IBM-3} is rather complex
and as yet their analysis is incomplete.
The cases with dynamical U(6) symmetry [or SU(3) charge symmetry]
were studied in detail in reference~\cite{Garcia99}.
Other classifications that conserve $J$ and $T$ [but not charge SU(3)]
were proposed and analyzed in references~\cite{Ginocchio96,Kota98}.

All bosons included in \mbox{IBM-3} have $T=1$
and, in principle, other bosons can be introduced
that correspond to $T=0$ neutron--proton pairs.
This further extension
(proposed by Elliott and Evans~\cite{Elliott81} and referred to as \mbox{\bf IBM-4})
can be considered as the most elaborate version of the IBM.
There are several reasons for including also $T=0$ bosons.
One justification is found
in the $LS$-coupling limit of the nuclear shell model,
where the two-particle states of lowest energy
have orbital angular momenta $L=0$ and $L=2$
with $(S,T)=(0,1)$ or (1,0).
Furthermore, the choice of bosons in \mbox{IBM-4}
allows a boson classification
containing Wigner's supermultiplet algebra SU(4).
These qualitative arguments in favour of \mbox{IBM-4}
have been corroborated by quantitative, microscopic studies
in even--even~\cite{Halse84}
and odd--odd~\cite{Halse85} $sd$-shell nuclei.

Arguably the most important virtue
of the extended versions \mbox{IBM-3} and \mbox{IBM-4}
is that they allow the construction of dynamical symmetries in the IBM
with quantum numbers that have their counterparts in the shell model
(isospin, Wigner supermultiplet labels, {\it etc.}).
As so often emphasized by Elliott~\cite{Elliott85},
this feature allows tests of the validity of the IBM
in terms of the shell model.

\subsection{Supersymmetry}
\label{ss_susy}
Symmetry techniques can be applied to systems of interacting bosons
and to systems of interacting fermions.
In both cases the dynamical algebra is ${\rm U}(\Omega)$,
with $\Omega$ the number of states available to a single particle.
In both cases solvable models can be constructed
from the study of the subalgebras of ${\rm U}(\Omega)$.
Not surprisingly, the same symmetry techniques
can be applied to systems composed of interacting bosons {\em and} fermions.
If the bosons and fermions commute,
the dynamical algebra of the boson--fermion system is
${\rm U}^{\rm B}(\Omega_{\rm b})\otimes{\rm U}^{\rm F}(\Omega_{\rm f})$,
and the study of its subalgebras again leads to solvable hamiltonians.

This idea was applied in the context of the interacting boson--fermion model ({\bf IBFM})
which proposes a description of odd-mass nuclei
by coupling a fermion to a bosonic core~\cite{Iachello79}.
Properties of even--even and odd-mass nuclei can be obtained
from IBM and IBFM, respectively,
but no unified description is achieved with the dynamical algebra 
${\rm U}^{\rm B}(6)\otimes{\rm U}^{\rm F}(\Omega)$
which does not contain both types of nuclei
in a single of its irreducible representations.
Nuclear supersymmetry provides a theoretical framework
where bosonic and fermionic systems are treated
as members of the same {\bf supermultiplet}
and where excitation spectra of the different nuclei
arise from a single hamiltonian.
A necessary condition for such an approach to be successful
is that the energy scale
for bosonic and fermionic excitations is comparable
which is indeed the case in nuclei.
Nuclear supersymmetry was originally postulated
by Iachello and co-workers~\cite{Iachello80,Balantekin81,Balantekin82,Bijker84}
as a symmetry among doublets and was subsequently extended to quartets of nuclei
which include an odd--odd member~\cite{Isacker85}.

Schematically, states in even--even and odd-mass nuclei
are connected by the generators
\begin{equation*}
\left(\begin{array}{c|c}
b^\dag b&0\\
---&---\\
0&a^\dag a
\end{array}\right),
\end{equation*}
where $a$ ($b$) refers to a fermion (boson) and indices are omitted for simplicity.
States in an even--even nucleus are connected by the operators in the upper left-hand corner
while those in odd-mass nuclei require both sets of generators.
No operator connects even--even to odd-mass states.
An extension of this algebraic structure considers in addition operators
that transform a boson into a fermion or {\it vice versa},
\begin{equation*}
\left(\begin{array}{c|c}
b^\dag b&b^\dag a\\
---&---\\
a^\dag b&a^\dag a
\end{array}\right).
\end{equation*}
This set does not any longer form a {\sl classical Lie algebra}
which is defined in terms of commutation relations.
For example,
\begin{equation*}
[a^\dag b,b^\dag a]=
a^\dag bb^\dag a-b^\dag aa^\dag b=
a^\dag a-b^\dag b+2b^\dag ba^\dag a,
\end{equation*}
which does not close into the original set
$\{a^\dag a,b^\dag b,a^\dag b,b^\dag a\}$.
The inclusion of the cross terms does not lead to a classical Lie algebra
since the bilinear operators $b^\dag a$ and $a^\dag b$
do not behave like bosons but rather as fermions,
in contrast to $a^\dag a$ and $b^\dag b$,
both of which have bosonic character.
This suggests the separation of the generators in two sectors,
the bosonic sector $\{a^\dag a,b^\dag b\}$
and the fermionic sector $\{a^\dag b,b^\dag a\}$.
Closure is maintained by considering {\em anti-}commutators among the latter operators
and commutators otherwise.
This leads to the {\bf graded} or {\bf superalgebra} is ${\rm U}(6/\Omega)$,
where 6 and $\Omega$ are the dimensions of the boson and fermion algebras.

By embedding ${\rm U}^{\rm B}(6)\otimes{\rm U}^{\rm F}(\Omega)$
into a superalgebra ${\rm U}(6/\Omega)$,
the unification of the description
of even--even and odd-mass nuclei is achieved.
Formally, this can be seen from the reduction
\begin{equation*}
\begin{array}{ccccc}
{\rm U}(6/\Omega)&\supset&
{\rm U}^{\rm B}(6)&\otimes&{\rm U}^{\rm F}(\Omega)\\
\downarrow&&\downarrow&&\downarrow\\[0mm]
[{\cal N}\}&&[N_{\rm b}]&&[1^{N_{\rm f}}]
\end{array}.
\end{equation*}
The supersymmetric irreducible representation $[{\cal N}\}$
of ${\rm U}(6/\Omega)$
imposes symmetry in the bosons
and anti-symmetry in the fermions,
and contains the ${\rm U}^{\rm B}(6)\otimes{\rm U}^{\rm F}(\Omega)$
irreducible representations $[N_{\rm b}]\times[1^{N_{\rm f}}]$
with ${\cal N}=N_{\rm b}+N_{\rm f}$~\cite{Balantekin81}.
A single supersymmetric irreducible representation
therefore contains states in even--even ($N_{\rm f}=0$) as well as odd-mass ($N_{\rm f}=1$) nuclei.

\begin{figure}
\centering
\includegraphics[width=15cm]{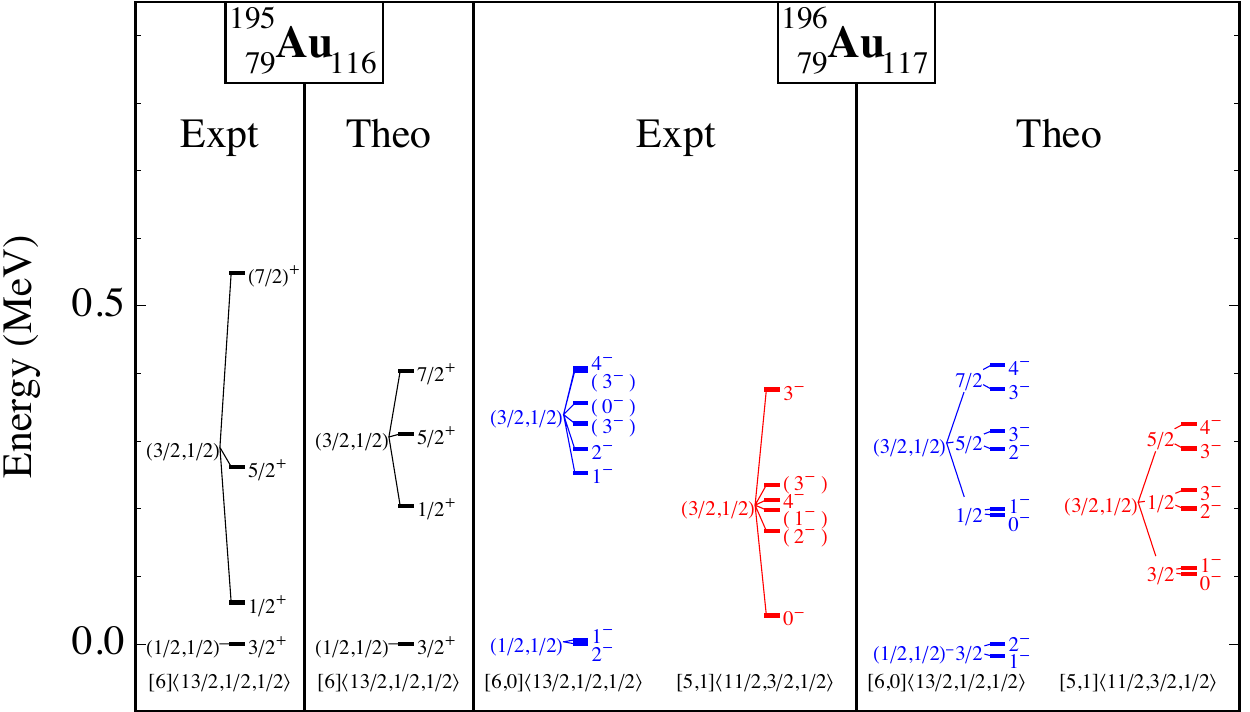}
\includegraphics[width=15cm]{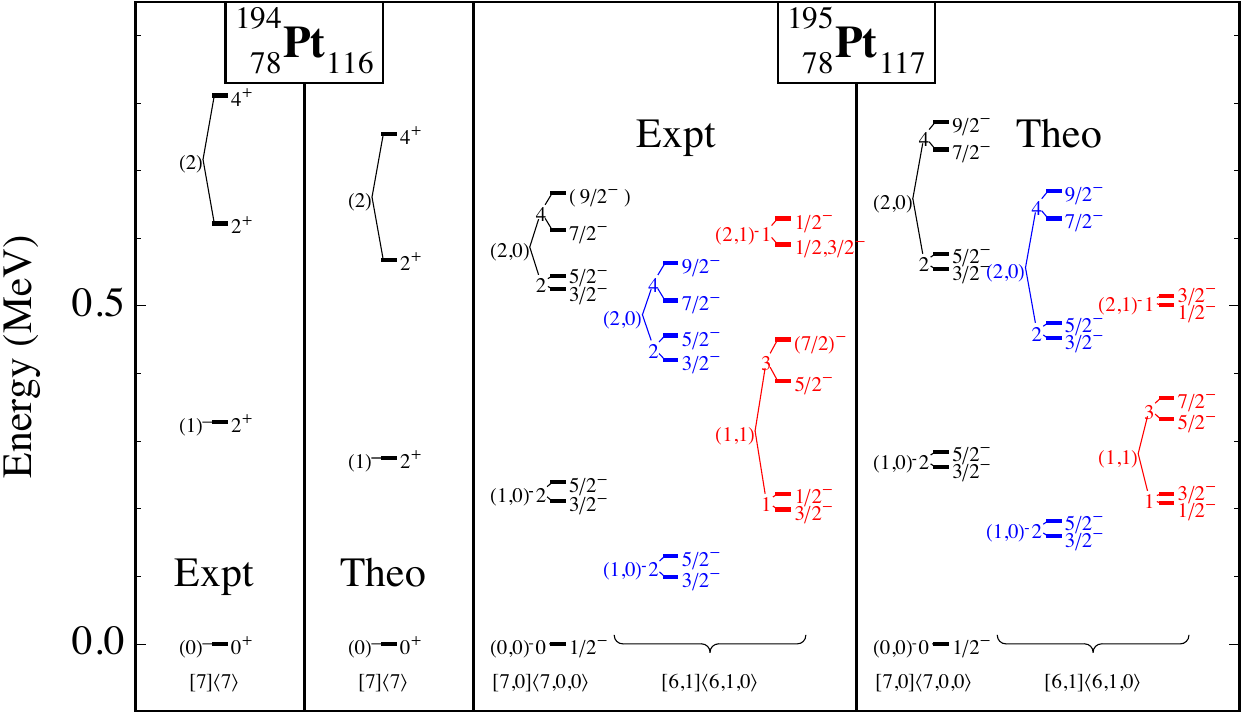}
\caption{
Example of a ${\rm U}_\nu(6/12)\otimes{\rm U}_\pi(6/4)$ supermultiplet.}
\label{f_quartet}
\end{figure}

Finally, if a distinction is made between neutrons and protons,
it is natural to propose a generalized dynamical algebra
${\rm U}_\nu(6/\Omega_\nu)\otimes{\rm U}_\pi(6/\Omega_\pi)$
where $\Omega_\nu$ and $\Omega_\pi$
are the dimensions
of the neutron and proton single-particle spaces, respectively.
This algebra contains generators
which transform bosons into fermions and {\it vice versa},
and furthermore are distinct for neutrons and protons.
The supermultiplet now contains
a quartet of nuclei (even--even, even--odd, odd--even and odd--odd)
which are to be described simultaneously with a single hamiltonian.
The predictions of ${\rm U}_\nu(6/12)\otimes{\rm U}_\pi(6/4)$
have been extensively investigated
in platinum ($Z=78$) and gold ($Z=79$) nuclei,
where the dominant orbits are
$3p_{1/2}$, $3p_{3/2}$ and $2f_{5/2}$ for the neutrons,
and $2d_{3/2}$ for the protons.
Probing the properties of the odd--odd member of the quartet
proved to be a challenge
and it took many years of dedicated experiments
to establish a convincing case of a complete supermultiplet~\cite{Metz99}
which is shown in figure~\ref{f_quartet}.

%{\bf In the present paper, we opted to study exactly solvable models that single
%out the most important degrees of freedom as observed in atomic nuclei, such
%as the presence of nuclear superfluidity, the characteristic short-range characteristic
%generic to all effective two-nucleon effective interactions near closed shells and  
%the appearance of nuclear collective modes of motion (vibrations, rotations,..).
%Thereby, we approximate the more complex microscopic theoretical approaches (the 
%nuclear shell model, including general NN interactions, (beyound) self-consistent
%mean-field methods) but keeping the essential phyiscs ingredients. Still, the results
%deriving from a study of exactly solvable models can be used as reference
%results when comparing with nuclear data over large regions (series of isotopes or
%isotone) and comparing with the most advanced numerical approaches used in describing
%nuclear structure properties.}   

\section{Beyond exact solvability}
\label{s_bes}
The exact solutions discussed in this review
are restricted to particular hamiltonians
of the nuclear shell model, the geometric collective model
and the interacting boson model.
This concluding section contains a succinct and qualitative discussion
of model hamiltonians that are not exactly solvable for all eigenstates
but only for a subset of them.

It is well known that only a limited number of potentials in quantum mechanics
are analytically solvable,
meaning that the entire energy spectrum
of eigenvalues and corresponding eigenfunctions
can be obtained as exact solutions.
A wider class of potentials can be constructed,
with an exact solution for a finite (or possibly infinite) but not complete
part of the eigenvalue spectrum.
Models with such potentials are called quasi-exactly solvable (QES).
This is a rich field of research
that has been studied since many years
(see, {\it e.g.}, reference~\cite{Ushveridze94} and references therein).
Very few QES applications were considered up to now in nuclear structure,
one of which was cited in the context of the Bohr hamiltonian~\cite{Levai04}.

A related generalization concerns dynamical symmetries.
The conditions for a dynamical symmetry
are seldom satisfied in the description of complex quantum many-body systems.
A more realistic description requires the breaking of the dynamical symmetry
by adding, in a particular subalgebra chain, one or more terms from a different chain.
This, in general, results in the loss of complete solvability.
Nevertheless, hamiltonians with a partial dynamical symmetry (PDS) can be constructed,
such that a subset of its eigenstates is characterized
by a subset of the labels of a particular dynamical symmetry.
The generic mechanism is layed out precisely by Alhassid and Leviatan~\cite{Alhassid92}
and extensively discussed in the review of Leviatan~\cite{Leviatan11}.
Three types of PDS exist
depending on whether all (or part) of the eigenstates
carry all (or part) of the quantum numbers
associated with the dynamical symmetry.   

Many nuclei can be described
as exhibiting a transition between two dynamical symmetries
({\it e.g.}, in IBM from U(5) to SU(3) or from U(5) to O(6),
or a transition from pairing SU(2) to rotor SU(3), {\it etc.}).
Although the transitional hamiltonian
in general does not have a dynamical symmetry,
it turns out that, except for a very narrow region before (or after) the transition point,
the initial (or final) symmetry remains intact in some effective way.
This is possible because of the existence
of a quasi-dynamical symmetry (QDS)~\cite{Rowe04b,Turner05,Rosensteel05},
formulated in a precise way by Rowe {\it et al.}
using the concept of embedded representations~\cite{Rowe88}.
Strictly speaking a hamiltonian with QDS is not exactly solvable.
However, the concept of QDS
clearly emanates from that of dynamical symmetry,
with applications in the study of atomic nuclei~\cite{Cejnar10}
and of more general systems~\cite{Rowe12}.

\section*{Further reading}
Scientific studies covering a period of almost 80 years
are difficult to summarize in barely 30 pages
and consequently most developments were only fleetingly discussed in the present review.
It is therefore appropriate to end with a list of suggestions for further reading.
Many books exist on symmetries in physics and group theory.
A standard monograph is the one of Hamermesh~\cite{Hamermesh62};
a more recent one in the spirit of this review is by Iachello~\cite{Iachello06}.
Nuclear structure is comprehensively covered
in the standard works by Bohr and Mottelson~\cite{Bohr69,Bohr75}
and the many-body techniques used in the field
are discussed by Ring and Schuck~\cite{Ring80}.
Details on the shell model can be found in references~\cite{Heyde90,Talmi93}
while the interacting boson model is covered in references~\cite{Talmi93,Iachello87,Iachello91}.
A recent monograph~\cite{Frank09}
gives an overview of symmetries encountered in the description of atomic nuclei.
Finally, a discussion on embedding algebraic collective models
within a shell-model framework
can be found in the book of Rowe and Wood~\cite{Rowe10}.

\end{document}